\documentclass[aps,prd,superscriptaddress,nofootinbib,amsfonts,amssymb,amsmath,notitlepage]{revtex4-1}

\usepackage{graphicx}
\usepackage{float}
\usepackage{wrapfig}
\usepackage{bm}
\usepackage{color}
\usepackage{xcolor}
\usepackage{dcolumn}
\usepackage{booktabs}
\usepackage{physics}
\usepackage{tensor}
\usepackage{empheq}
\usepackage{amsmath}
\usepackage{amssymb} 
\usepackage{hyperref}
\hypersetup{
colorlinks=true,
citecolor=blue,
citebordercolor=red,
linktoc=all,
linkcolor=blue,
urlcolor=blue
}
\usepackage[noabbrev]{cleveref}
\usepackage{enumitem}

\catcode`\@=11
\def\@versim#1#2{\vcenter{\offinterlineskip
\ialign{$\m@th#1\hfil##\hfil$\crcr#2\crcr\sim\crcr } }}
\catcode`\@=12

\newcommand{\gsim}{\gtrsim}
 
\newcommand{\bvec}[1]{\mbox{\boldmath $#1$}}

\newcommand{\p}{\partial}

\newcommand{\bp}{\begin{pmatrix}}
\newcommand{\ep}{\end{pmatrix}}
\newcommand{\nn}{\nonumber\\}

\newcommand{\del}{\partial}

\newcommand{\df}{\text{d}}

\newcommand{\bs}[1]{\boldsymbol}

\DeclareSymbolFont{symbolsC}{U}{pxsyc}{m}{n}

\def\coloneqq{\mathrel{\mathop:}=}

\newcommand{\mc}[1]{\mathcal{#1}}
\newcommand{\sfdv}[3]{\frac{\delta^2 #1}{\delta #2 \delta #3}}

\makeatother
\graphicspath{{./Figs/}}

\begin{document}


\newbox{\ORCIDicon}
\sbox{\ORCIDicon}{\large
                  \includegraphics[width=0.8em]{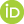}}

\title{
Gradient Flow Exact Renormalization Group for Scalar Quantum Electrodynamics
}

\author{Junichi  \surname{Haruna}\,\href{https://orcid.org/0000-0002-1828-8183}{\usebox{\ORCIDicon}}}
\email{j.haruna111@gmail.com}
\affiliation{Center for Quantum Information and Quantum Biology, Osaka University, Toyonaka, Osaka 560-0043, Japan}

\author{Masatoshi \surname{Yamada}\,\href{https://orcid.org/0000-0002-1013-8631}{\usebox{\ORCIDicon}}}
\email{yamada@jlu.edu.cn}
\affiliation{Center for Theoretical Physics and College of Physics, Jilin University, Changchun 130012, China}

\begin{abstract}
Gradient Flow Exact Renormalization Group (GF-ERG) is a framework to define the renormalization group flow of Wilsonian effective action utilizing coarse-graining along the diffusion equations.
We apply it for Scalar Quantum Electrodynamics and derive flow equations for the Wilsonian effective action with the perturbative expansion in the gauge coupling.
We focus on the quantum corrections to the correlation functions up to the second order of the gauge coupling and discuss the gauge invariance of the GF-ERG flow.
We demonstrate that the anomalous dimension of the gauge field agrees with the standard perturbative computation and that the mass of the photon keeps vanishing in general spacetime dimensions.
The latter is a noteworthy fact that contrasts with the conventional Exact Renormalization Group formalism in which an artificial photon mass proportional to a cutoff scale is induced.
Our results imply that the GF-ERG can give a gauge-invariant renormalization group flow in a non-perturbative way.
\end{abstract}

\maketitle

\section{Introduction}
The Exact Renormalization Group (ERG)~\cite{Wilson:1973jj,Polchinski:1983gv,Wegner:1972ih,Wetterich:1992yh} is a powerful tool for addressing non-perturbative phenomena in quantum field theory~\cite{Morris:1993qb,Morris:1998da,Berges:2000ew,Aoki:2000wm,Bagnuls:2000ae,Polonyi:2001se,Pawlowski:2005xe,Gies:2006wv,Delamotte:2007pf, Rosten:2010vm,Kopietz:2010zz,Braun:2011pp,Dupuis:2020fhh}.
In particular, ERG techniques have been employed to explore critical phenomena and phase transitions in strongly correlated systems, such as the Hubbard model in condense-matter physics and quantum chromodynamics in high-energy physics.

The central idea of the ERG is the coarse-graining for quantum fluctuations within the path integral.
Integrating out quantum fluctuations with higher momentum $|p|\gsim \Lambda$ with an artificial cutoff $\Lambda$ embodies the coarse-graining.
This procedure entails the Wilsonian effective action $S_\Lambda$, which contains infinite effective operators.
In this framework, a functional differential equation describes the change of the effective action $S_\Lambda$ induced by quantum fluctuations under the cutoff variation~\cite{Wilson:1973jj,Polchinski:1983gv,Wegner:1972ih,Wetterich:1992yh} and is called ``the ERG equation".

However, the introduction of the explicit momentum cut-off conflictes with gauge symmetries.
Since preserving gauge symmetries in describing quantum systems is essential to compute physical quantities, using the ERG tends to be avoided, especially in particle physics.
Several attempts~\cite{Litim:1998wk,Litim:1998nf,Wetterich:2016ewc,Asnafi:2018pre,Igarashi:2019gkm,Igarashi:2021zml,Igarashi:2016gcf,Fejos:2016wza,Fejos:2017sjl} have been made to realize a gauge-invariant formulation of non-perturbative renormalization group methods. 
For instance, a manifestly gauge-invariant formulation for ERG in Yang-Mills theories has been developed~\cite{Morris:1999px,Morris:2000fs}.
This formulation successfully reproduces the perturbative two-loop beta function for the gauge coupling without requiring gauge-fixing~\cite{Morris:2005tv}.
Moreover, this approach has been employed in QED at both the perturbative one-loop level~\cite{Arnone:2005vd} and the non-perturbative level~\cite{Rosten:2008zp}, as well as in gravity~\cite{Morris:2016nda}.

Among them, the Gradient Flow Exact Renormalization Group (GF-ERG) is a potential framework of the gauge-invariant formulation of the ERG and has recently been suggested in Ref.~\cite{Sonoda:2020vut}.
One of its novel features is the incorporation of coarse-graining along the flow of diffusion equations (``gradient flow'') for field operators into the RG flow.
From the GF-ERG perspective, the conventional ERG formalism with the exponentially damping UV cutoff function is based on the simple diffusion equation, which does not respect gauge symmetries.
One can derive a new type of ERG equation with GF-ERG by replacing this simple diffusion equation with a general one.
A unique feature of the GF-ERG formalism is that it can define RG flows that inherit local or global symmetries of the general diffusion equations.
Leveraging this attribute allows us to define manifestly gauge-invariant renormalization group flows through GF-ERG.

GF-ERG has been applied in various physical systems, such as pure Yang-Mills theory~\cite{Sonoda:2020vut}, quantum electrodynamics (QED)~\cite{Miyakawa:2021hcx, Miyakawa:2021wus}, scalar field theory~\cite{Abe:2022smm}, and free Dirac fermions with background gauge fields~\cite{Miyakawa:2023yob}.
These studies show that the GF-ERG successfully realizes the invariance of symmetries along renormalization group flows, at least at the perturbative level.
In particular, Refs.~\cite{Miyakawa:2021hcx, Miyakawa:2021wus} have studied QED (describing the interaction between the electrons and the photon) and have shown that the gauge invariance is maintained in the sense that the two-point correlation function of the gauge field satisfies transversality.
Furthermore, the counterpart of the ERG equation in the GF-ERG formalism has been derived and called ``the GF-ERG equation''.
It is a highly non-trivial extension of the ERG equation and involves additional loop corrections from modifying diffusion equations.

Here, we are interested to know whether the GF-ERG formulation realizes the gauge invariance in other systems.
Our work aims to explore this point by examining a well-established physical system, namely scalar quantum electrodynamics (sQED). 
In this paper, we show that the GF-ERG equation yields the transverse kinetic term of the gauge field, that is, $2\gamma_1 (k^2 - k^\mu k^\nu)$ with an appropriate anomalous dimension at the perturbative one-loop level $\gamma_1=1/48\pi^2$ in four dimensions.
Furthermore, we also show that the mass of the gauge field is not induced at one-loop level in general dimensions.
The latter result contrasts with the conventional ERG equation, which yields a finite gauge field mass proportional to the cutoff scale.
These findings indicate that GF-ERG could realize a gauge-invariant formulation for the ERG.

In this study, we work on the following procedures: First, we define the Wilsonian effective action for sQED utilizing GF-ERG and investigate the condition of the BRST invariance of the effective action (or Ward-Takahashi identity).
Second, we derive the GF-ERG equations for vertices in the Wilsonian effective action order by order with perturbative expansions in the sQED gauge coupling.
Finally, we construct the two-point correlation function of the gauge field from which the anomalous dimension and loop corrections to the mass of the gauge field are read off.
In particular, we show that the anomalous dimension agrees with the standard perturbative calculation in four dimensions and that the mass of the gauge field does not receive loop corrections in general dimensions.

This paper is organized as follows.
In \Cref{sec: conventional Wilson-Polchinski equation}, we briefly review the conventional formulation of ERG.
We also give an alternative definition of the Wilsonian effective action, leading to the idea of GF-ERG.
In \Cref{sec: GF-ERG for sQED}, we define the Wilsonian effective action for sQED, combining GF-ERG and a set of diffusion equations with BRST invariance.
Then we derive the flow equation of the effective action (``GF-ERG equation'') and the Ward-Takahashi identity, which is the result of the manifest gauge invariance [or Becchi-Rouet-Stora-Tyutin (BRST) invariance] of the GF-ERG flow.
In \Cref{sec: Perturbative analysis of GF-ERGeq}, we solve the GF-ERG equation for sQED perturbatively up to the second order in the gauge coupling.
We investigate the one-loop correction to the two-point function of the $U(1)$ gauge field to see that the mass term of the gauge field is not generated in the second-order correction from the matter field.
\Cref{sec: conclusions} is devoted to conclusions and discussions.

\section{Wilson-Polchinski equation and diffusion equation}
\label{sec: conventional Wilson-Polchinski equation}
The renormalization group equation is expressed as a functional differential equation whose form is not unique in quantum field theory.
In this work, we deal with the Wilson-Polchinski-type equation.
This section briefly reviews the conventional Wilson-Polchinski equation~\cite{Wilson:1973jj,Polchinski:1983gv} to highlight a difference from the GF-ERG equation.

Let us start with the path integral for a scalar field theory described by the bare action $S_{0}[\phi]$ defined on a certain energy scale $\Lambda_0$ where the field $\phi$ is a function of momentum $p$.  More specifically, the action at $\Lambda_0$ is given by
\begin{align}
 S_0[\phi]= -\frac{1}{2}\int_p\left[\frac{p^2}{K(p/\Lambda_0)} \phi(-p) \phi(p) \right] + S_0^I[\phi]\,,
\end{align}
where $S_0^I[\phi]$ is the interaction part of the action and $\int_p\coloneqq\int \df^D p/(2\pi)^D$ is a short-hand notation for the momentum integral in $D$ dimensional spacetime. Here, $K(p/\Lambda_0)$ is the regulator as a smooth function of $p/\Lambda_0$ and behaves as
\begin{align}
K(p/\Lambda_0)= \begin{cases}
1 & \text{for $p\to 0$}\,,\\
0 & \text{for $p\to \infty$}\,.
\end{cases}
\label{eq: K cutoff function}
\end{align}
This implies that the propagator of $\phi$ reads $K(p/\Lambda_0)/p^2$ and thus the theory is regularized in sense that the scalar field for $p>\Lambda_0$ does not propagate.

We now divide the momentum into the higher momentum mode $\Lambda \leq |p|\leq \Lambda_0$ and the lower one $|p|\leq \Lambda$ by introducing a certain cutoff scale $\Lambda$.
This can be implemented by the division of the field such that $\phi(p)=\phi(p) + \varphi(p)$ where
\begin{align}
\phi(p) = \begin{cases}
\phi(p) & \text{$p\leq \Lambda$, otherwise zero}; \\[1ex]
\varphi(p)& \text{$p>\Lambda$, otherwise zero}.
\end{cases}
\label{eq: field divided}
\end{align}
In the path integral formalism, the path integral measure is divided into $\mathcal D\phi=\mathcal D\phi \mathcal D\varphi$.
Then, by integrating out only the higher momentum mode field $\varphi$, we can defines the Wilsonian effective action $S_{\tau}[\phi]$ as
\begin{align}
e^{S_\tau[\phi]} = \int  {\mathcal D}\varphi\, e^{S_{0}[\phi+ \varphi]}  \,.
\end{align}
Here, we have introduced the dimensionless scale $\tau - \tau_0= \log(\Lambda_0/\Lambda)$ so that $S_{\tau_0}=S_0$. 

The Wilson-Polchinski equation describes the change of the Wilsonian effective action $S_\tau$ under varying the dimensionless scale $\tau$ (or equivalently $\Lambda$) in a functional differential equation.
We hand over its detailed derivation to, e.g. Refs.~\cite{Wilson:1973jj,Polchinski:1983gv,Dutta:2020vqo} and show the explicit form of the conventional Wilson-Polchinski equation here.
For our purposes, we first define dimensionless quantities with the dimensionless cutoff scale $\Lambda$ such that
\begin{align}
&\tilde\phi(\tilde p) = \Lambda^{d_\phi-D} \phi(p)\,,&
&\tilde p= p/\Lambda\,,
\label{eq: making dimensionless}
\end{align}
with $d_\phi$ the mass-dimension of the field $\phi$.
With these quantities, the Wilson-Polchinski equation is given by
\begin{align}
 \partial_\tau e^{S_\tau[\tilde\phi]}&=\int_{\tilde p} {\left[\frac{\Delta(\tilde p)}{K(\tilde p)} -( d_\phi -D)  -\gamma_\phi+\tilde p \cdot \p_{{\tilde p}}\right] \tilde\phi(\tilde p) \cdot \frac{\delta}{\delta \tilde\phi(\tilde p)} e^{S_\tau[\tilde\phi]} } \nn
& \quad
+\int_{\tilde p}\left\{\frac{1}{\tilde p^2}\left[\frac{2 \Delta(\tilde p) {\mathcal K}(\tilde p)}{K(\tilde p)}+\tilde p \cdot \p_{{\tilde p}} {\mathcal K}(\tilde p)\right]-2 \gamma_\phi\right\} \frac{1}{2} \frac{\delta^2}{\delta \tilde\phi(\tilde p) \delta \tilde\phi(-\tilde p)} e^{ S_\tau[\tilde\phi]}\, ,
\label{eq: dimensionless Wilson-Polchinski equation}
\end{align}
where we have introduced short-hand notations, $\tilde p\cdot \del_{\tilde p} = \tilde p_\mu{\del}/{\del \tilde p_\mu}$ and 
\begin{align}
&\Delta(\tilde p) = \Lambda\frac{\partial}{\partial \Lambda}K(\tilde p)\,,&
&{\mathcal K}(\tilde p) = K(\tilde p)\Big(1-K(\tilde p)\Big) \,.
\label{eq: cutoff function mathcal K}
\end{align}
and $\gamma_\phi$ is the anomalous dimension of $\phi$.

\begin{figure}
\centering
\includegraphics[width=0.7\linewidth,pagebox=cropbox,clip]{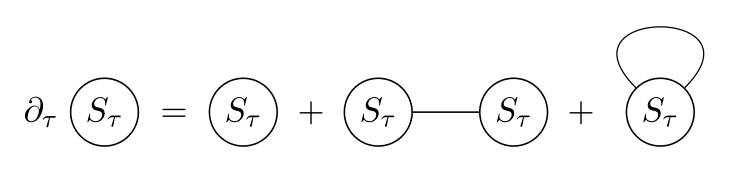}
\caption{Diagrammatic representation of the conventional Wilson-Polchinski equation \eqref{eq: dimensionless Wilson-Polchinski equation}. The solid line denotes the propagator of the field $\phi$.
The first term on the right-hand side corresponds to the canonical scaling due to the introduction of dimensionless quantities.
Thus, the second term (the dumbbell diagram) and third term (the ring diagram) corresponds to quantum corrections.
The solid line depicts the (modified) propagator.}
\label{fig:The diagrammatic representation of the WP equation}
\end{figure}

Hereafter, we deal with the dimensionless version of the Wilson-Polchinski equation and, therefore, remove tildes on dimensionless quantities.
The Wilson-Polchinski equation \eqref{eq: dimensionless Wilson-Polchinski equation} for the Wilsonian effective action $S_\tau$ is represented schematically in \Cref{fig:The diagrammatic representation of the WP equation}. The first term on the right-hand side in \Cref{eq: dimensionless Wilson-Polchinski equation} is the canonical scaling, while the second term induces quantum corrections.
The second-order functional derivative acting on $e^{ S_\tau[\phi]}$ yields
\begin{align}
\frac{\delta^2}{\delta \phi( p) \delta \phi(- p)} e^{ S_\tau[\phi]} =
 e^{ S_\tau[\phi]}  \left[ \frac{\delta S_\tau[\phi]}{\delta \phi( p)} \frac{\delta S_\tau[\phi]}{\delta \phi(- p)} + \frac{\delta^2 S_\tau[\phi]}{\delta \phi( p) \delta \phi(- p)}  \right]\,.
\end{align}
Thus, the first term in the square bracket on the right-hand side of this equation corresponds to the dumbbell diagram in \Cref{fig:The diagrammatic representation of the WP equation}, while the second term is represented by the ring (loop) diagram.
The solid line shows the (modified) propagator given in the bracket $\{~\}$ of the second term in \Cref{eq: dimensionless Wilson-Polchinski equation}.

Now, we intend to formally solve the Wilson-Polchinski equation \eqref{eq: dimensionless Wilson-Polchinski equation}. To this end, let us take the following cutoff function that satisfies Eq.~\eqref{eq: K cutoff function}: 
\begin{align}
K(p) = e^{-p^2}\,.
\label{eq: cutoff function K}
\end{align}
For this choice, the formal solution to the Wilson-Polchinski equation \eqref{eq: dimensionless Wilson-Polchinski equation} is found to be~\cite{Sonoda:2020vut}
\begin{align}
e^{S_{\tau}[\phi]}
\coloneqq \hat{s}^{-1} \int {\mathcal D}\phi \prod_{x,\mu}
\delta\qty(\phi(x) - e^{-\Theta_\phi(\tau)} \phi'(t,e^{\tau-\tau_0}x)) 
\hat{s}' e^{S_{\tau_0}[\phi']}\,.
\label{eq: GF-ERG equation}
\end{align}
Here, we have introduced the operator
\begin{align}
\hat{s} \coloneqq \exp(-\frac{1}{2} \int_p \frac{\mathcal K(p)}{p^2} \frac{\delta^2}{\delta\phi^*(p)\delta\phi(p)})\,,
\label{eq: scrambler original}
\end{align}
which is called the ``scrambler", and the scaling factor of $\phi$,
\begin{align}
\Theta_\phi(\tau)=\int^\tau_{\tau_0}d\tau' \qty(d_\phi+\gamma_\phi({\tau'}))\,.
\end{align}

\Cref{eq: GF-ERG equation} is can be interpreted as a renormalization transformation from the initial action $S_{\tau_0}[\phi']$ to $S_\tau[\phi]$. This renormalization transformation naturally entails the flowing field $\phi'(t, e^{\tau-\tau_0}x)$ with $t-t_0=e^{2(\tau - \tau_0)}-1$, as a solution to the diffusion equation
\begin{align}
\label{eq: simple diffusion equation}
\del_t \phi'(t,x)=\del^2 \phi'(t,x)\,,
\end{align}
with the initial conditions at $t=t_0$ (or equivalently $\tau=\tau_0$)
\begin{align}
\phi'(t=t_0,x) =\phi'(x)\,.
\end{align}
Therefore, the conventional Wilson-Polchinski equation already involves information on the gradient flow of the field \eqref{eq: simple diffusion equation}.
In this sense, GF-ERG can be regarded as an extension of ERG.

\section{GF-ERG for Scalar QED}
\label{sec: GF-ERG for sQED}
In this section, we formulate GF-ERG for sQED following Ref.~\cite{Sonoda:2020vut}. In \Cref{sec: Definition of Wilsonian effective action}, we introduce the $U(1)$ gauge field, the (anti-)ghost, Nakanishi-Lautrup field and matter scalar field and discuss their diffusion equations which are consistent with the BRST transformation.
The right-hand side of \cref{eq: simple diffusion equation} is not invariant for the gauge (or BRST) transformation, and consequently the conventional Wilson-Polchinski equation is not a gauge-invariant formulation.
In the general GF-ERG approach, we deform the simple diffusion equation \eqref{eq: simple diffusion equation} so that the diffusion equations respect the BRST symmetry.
Then, we define the Wilsonian effective action based on these diffusion equations with GF-ERG.
In \Cref{sec: GF-ERGE equation for sQED}, we derive the GF-ERG equation of this Wilsonian effective action, taking the derivative of it with respect to the scale $\tau$.
In \Cref{sec: Tree level action}, we study the Gaussian fixed-point action of the GF-ERG equation, which will be used later in performing the perturbative analysis in \Cref{sec: Perturbative analysis of GF-ERGeq}.
Finally, we investigate the (modified) BRST invariance of the GF-ERG equation and derive the Ward-Takahashi identity for the Wilsonian effective action in \Cref{sec: Ward-Takahashi identity}.

\subsection{Definition of Wilsonian effective action}
\label{sec: Definition of Wilsonian effective action}
As we have mentioned in \Cref{sec: conventional Wilson-Polchinski equation}, the renormalization transformation for the Wilsonian effective action from $S_{\tau_0}$ to $S_{\tau}$ can be expressed as
\begin{align}
e^{S_{\tau}[\psi]}
= \hat{s}^{-1} \int {\mathcal D}\psi \prod_{x,\mu}
\delta\qty(\psi(x) - e^{-\Theta_\psi(\tau)} \psi'(t,e^{\tau-\tau_0}x)) 
\hat{s}' e^{S_{\tau_0}[\psi']}\,.
\label{eq: renormalization transformation for sQED}
\end{align}
Note that all quantities and variables are supposed to be dimensionless thanks to the multiplication of an appropriate power of the energy scale $\Lambda$ 
(See \cref{eq: making dimensionless}).
Here, $S_{\tau}$ (and $S_{\tau_0}$) contains the interactions of the ghost fields $c$, $\bar c$ and the Nakanishi-Lautrup field $B$ in addition to the $U(1)$ gauge field $A_\mu$ and the complex scalar field $\phi$.
For a short-hand notation, we denote the superfield by $\psi$ as $\psi=(A_\mu, c, \bar c, B,\phi, \phi^*)$ in the present system, in which the canonical mass dimensions are given by
\begin{align}
&d_A= \frac{D-2}{2}\,,&
&d_B=\frac{D}{2}\,,&
&d_c=\frac{D-4}{2}\,,&
&d_{\bar c} =\frac{D}{2}\,,&
&d_\phi = d_{\phi^*} = \frac{D-2}{2}\,,
\label{eq: canonical mass dimensions of fields}
\end{align}
Note that the mass dimensions of the ghost fields are determined from $[\int\df^D x\, \bar c\, \p_\mu D^\mu c]=0$, i.e. $[\bar c c]=D-2$ where $[O]$ denotes the mass dimension of the physical quantity $O$.
Since the anti-ghost field $\bar c$ is {\it not} the complex conjugation of the ghost field $c$, but is an independent field of $c$, its mass dimension can be assigned independently. As will be seen later, the assignment for $d_c$ and $d_{\bar c}$ in \Cref{eq: canonical mass dimensions of fields} is fixed by consistency with the BRST transformation. For the same reason, it turns out that we can set $\gamma_A=\gamma_c= -\gamma_{\bar c}=-\gamma_B$ and $\gamma_\phi=\gamma_{\phi^*}$. The scrambler operator $\hat{s}$ in this system is given by
\begin{align}
\hat{s} \coloneqq \exp(\int_p \left[-\frac{1}{2} \fdv[2]{A_\mu(p)} + \frac{1}{2} \fdv[2]{B(p)}+  \fdv{c(p)}\fdv{\bar{c}(p)} - \fdv{\phi^*(p)}\fdv{\phi(p)} \right])\,.
\label{eq: scrambler}
\end{align}
Here, following Ref.~\cite{Sonoda:2020vut}, we have taken
\begin{align}
\label{e:ChoiceOfCalK}
    \mc{K}(p^2) = p^2\,.
\end{align}
This choice makes the derivation of the GF-ERG equation simple, but does not agree with the Wilson-Polchinski convention \eqref{eq: cutoff function mathcal K} with the choice of $K(p)$ as given in \Cref{eq: cutoff function K}.
The consistency of this choice of $\mc{K}(p^2)$ as a UV cutoff function will be discussed in \Cref{sec: Tree level action}.

The Wilsonian effective action $S_{\tau}$ obtained from \Cref{eq: renormalization transformation for sQED} with the simple diffusion equations \eqref{eq: simple diffusion equation} becomes equivalent to the solution to the conventional Wilson-Polchinski equation with cutoff functions \eqref{eq: cutoff function K} and \eqref{eq: cutoff function mathcal K}. However, these diffusion equations are not invariant under the following BRST transformation:\footnote{ 
The BRST invariance of the diffusion equations means that the flow along them are consistent with the BRST transformation.
More specifically, the relations like
\begin{align}
\del_t {\bvec \delta}_{\rm B} A_\mu = {\bvec \delta}_{\rm B} \del_t A_\mu
\end{align}
hold for all the fields $A_\mu, c, \bar{c}, B, \phi,$ and $\phi^*$.
}
\begin{align}
\label{eq: BRST transformation}
& {\bvec \delta}_{\rm B} A_\mu =  \del_\mu c\,,&
& {\bvec \delta}_{\rm B} c = 0\,,&
& {\bvec \delta}_{\rm B} \bar c =  B\,,&
& {\bvec \delta}_{\rm B} \phi = i e_0 c \phi\,,&
& {\bvec \delta}_{\rm B} B =0\,.
\end{align}
In this sense, the conventional Wilson-Polchinski equation is not BRST (gauge) invariant. Alternately, the BRST invariance under \Cref{eq: BRST transformation} suggests the modification of the diffusion equations for $\phi'$ and $\phi'^*$:
\begin{subequations}
\label{eq: GF diffusion equations}
\begin{align}
&\del_t \phi' (t,x) = D'{}^2 \phi'(t,x) + i e_0 \del_\mu A'_\mu (t,x) \phi'(t,x)\,,\\[1ex]
&\del_t {\phi'}^* (t,x) = {D'}^{*2} {\phi'}^*(t,x) - i e_0 \del_\mu A'_\mu (t,x) {\phi'}^*(t,x)\,,
\end{align}
\end{subequations}
where $D'_\mu \coloneqq \del_\mu - ie_0 A'_\mu(t,x)$.
The other fields obey the simple diffusion equations \eqref{eq: simple diffusion equation} in the same way as the conventional ERG.
It should be emphasized here that the above diffusion equations and the covariant derivative $D'_\mu$ are defined with the {\it bare} electric charge $e_0$.
This BRST invariance is inherited by (the flow of) the Wilsonian effective action, as will be seen in the next subsection.
The modified diffusion equations enforce the consistency with the BRST transformation on the fields at an arbitrary energy scale, and thus guarantee the BRST invariance of the system under the RG transformations.

Here, we explain the relation between the anomalous dimensions.
If the RG flow is invariant under the BRST transformation, the anomalous dimensions of $A_\mu$ and $c$ must agree with each other from \Cref{eq: BRST transformation}.
For the same reason, we find $\gamma_{\bar{c}} = \gamma_{\bar{B}}$.
Also, because the ghost sector in sQED is free,\footnote{
This discussion is valid, at least at the perturbative level.
Although it may not be so at the non-perturbative level, the framework of GF-ERG permits arbitrary assignment of anomalous dimensions for the (anti-)ghost.
Therefore, the assignment here should be regarded as an ansatz to derive consistent results to the perturbative theory, and GF-ERG itself can also be utilized in non-perturbative analyses.
} we expect that the ghost term does not receive the wave function renormalization.
Therefore, we find $\gamma_c = -\gamma_{\bar{c}}$, which leads to $\gamma_A = \gamma_c = -\gamma_{\bar{c}} = - \gamma_{B}$.
For the anomalous dimension of the matter field, the hermiticity of the action gives $\gamma_\phi = \gamma_{\phi^*}$.

\subsection{GF-ERG equation}
\label{sec: GF-ERGE equation for sQED}
Now, we derive the GF-ERG equation from \Cref{eq: renormalization transformation for sQED} and \Cref{eq: GF diffusion equations}. 
Differentiating the effective action $S_\tau$ with respect to $\tau$, we get
\begin{align}
  \partial_\tau e^{S_\tau}
&= (\text{R.H.S. of \cref{eq: dimensionless Wilson-Polchinski equation}})
  + \Bigg[ 4ie_\tau \int_x\,
  \widehat A_\mu(x) 
   \left( \partial_\mu +\frac{i}{2}e_\tau   \widehat A_\mu(x) \right)
   \fdv{\phi^*(x)} \left( \widehat {\phi}^*(x)e^{S_\tau} \right)
 + (\text{c.c}) \Bigg]\,.
\label{e:GF-ERGeq}
 \end{align}
Here, the first terms on the right-hand side are the same as the conventional Wilson-Polchinski equation \eqref{eq: dimensionless Wilson-Polchinski equation}, while the last term arises from the modification of the diffusion equations from \Cref{eq: simple diffusion equation} to \Cref{eq: GF diffusion equations}.
We have introduced the new field operators, $\widehat A_\mu(x)$, $\widehat {\phi}(x)$ and $\widehat {\phi}^*(x)$, which are obtained from the Fourier transformations for the field operators in momentum space:
\begin{align}
&\widehat A_\mu(p) = \hat s^{-1} A_\mu(p) \hat s =  A_\mu(p) + \frac{\delta}{\delta A_\mu(p)}\,,
\label{eq: hat fields}
\\
&\widehat {\phi}(p) =\hat s^{-1} \phi(p) \hat s= {\phi}(p)  + \frac{\delta}{\delta {\phi}^*(p) }\,,\\
&\widehat {\phi}^*(p) =\hat s^{-1} \phi^*(p) \hat s= {\phi}^*(p)  + \frac{\delta}{\delta {\phi}(p) }.
\end{align}
These terms contain the functional derivatives acting on $S_\tau$ and thus induce a number of interacting terms, e.g., $e^{-S_\tau}\widehat A_\mu e^{S_\tau} =  A_\mu + e_\tau \phi^*\del_\mu \phi +\cdots$. 
Since the complete GF-ERG equation \eqref{e:GF-ERGeq} is quite lengthy, it will be put in \Cref{app: GF-ERG}.
In the GF-ERG \cref{e:GF-ERGeq}, the renormalized electric charge $e_\tau$ is defined by
\begin{align}
\label{eq: definition of renormalized charge}
e_\tau & = e_0 \exp(-\int_{\tau_0}^\tau \df \tau'\, \qty(\frac{D-4}{2}+\gamma_A(\tau')))\,.
\end{align}
Whereas the conventional Wilson-Polchinski equation has tree- and one-loop structures, the GF-ERG contains higher loop terms originating from the modification of the diffusion equations.
In \Cref{sec: Perturbative analysis of GF-ERGeq}, we will derive the flow equations in terms of the perturbative expansion of $e_\tau$.

We note here that  the GF-ERG equation, in this way, may be regarded as a class of the Wegner equation~\cite{Wegner:1976bn,Latorre:2000qc},
\begin{align}
\label{eq: Wegner equation}
\p_\tau e^{S_\tau[\phi]} = \int_x \frac{\delta}{\delta \phi(x)}\left( \Psi(x) e^{S_\tau[\phi]} \right)\,.
\end{align}
Here, $\Psi(x)$ is the general ``coarse-graining operator". The conventional Wilson-Polchinski equation is reproduced by taking
\begin{align}
\Psi(x)e^{S_\tau[\phi]} = \int \mathcal D \phi\,\delta(\phi - b_\tau[\phi] )(\p_\tau b_\tau[\phi]) e^{S_0[\phi]}\,,
\end{align}
where $b_\tau[\phi]$ is the coarse-grained effective field as analogous to block-spin transformed variables in spin systems. In other words, this is the definition of $\Phi$ from $\Phi$ as given in \Cref{eq: field divided}.
For a general GF-ERG equation, one can derive the explicit form of the coarse-graining operator $\Psi(x)$; however, it is complicated, so we do not specify it here and discuss it in \Cref{app: GF-ERG}.

One of the main purposes of this study is to obtain the beta function by using the GF-ERG equation by means of perturbative expansion. 
To this end, we generalize $\phi\to \Phi=(A_\mu,B,\bar c,c, \phi, \phi^*)$, and divide the fields $\Phi$ into certain background fields $\bar\Phi$ and fluctuation fields $\Phi$ such that
\begin{align}
S_\tau [\Phi] = S^{(0)}[\bar\Phi] +  S^{(1)}[\bar\Phi] \Phi  + \frac{1}{2!}S^{(2)}[\bar\Phi] \Phi^2 + \frac{1}{3!}S^{(3)}[\bar\Phi] \Phi^3 + + \frac{1}{3!}S^{(4)}[\bar\Phi] \Phi^4  + \cdots\,,
\end{align}
where we have omitted the spacetime integrals.
The expansion allows us to write the GF-ERG equation \eqref{e:GF-ERGeq} in coupled functional differential equations for effective vertices $S^{(n)}[\bar\Phi]$ depending on the background field and external momenta.
The first term on the right-hand side provides the effective potential $V_{\rm ell}(\bar\Phi)$ for the background field $\bar\Phi$.
Therefore, the minimum of $V_{\rm ell}(\bar\Phi)$ determines the vacuum of the system. In this work, we assume that the symmetric vacuum, i.e. $\bar\Phi=0$, is realized.
The choice of an appropriate background (vacuum) should involve the equations of motion for the background fields $S^{(1)}[\bar\Phi]=0$, so that the second term vanishes.
In this work, we deal only with $S^{(2)}[\bar\Phi]$, $S^{(3)}[\bar\Phi]$ and $S^{(4)}[\bar\Phi]$ and truncate the higher-order terms $S^{(n)}[\bar\Phi]$ ($n\geq 5$). 
In the following, we construct the effective propagators $S^{(2)}[\bar\Phi]$ and effective vertices ($S^{(3)}[\bar\Phi]$, $S^{(4)}[\bar\Phi]$) by the GF-ERG.

\subsection{Construction of Gaussian part}
\label{sec: Tree level action}
To construct the Ward-Takahashi identity and perform the perturbative analysis of sQED with GF-ERG, we have to define the tree-level action, i.e. the Gaussian part of the action denoted by $\mc{S}_0^{(2)}$.
We briefly explain how to obtain the Gaussian action in the following.
First, we assume that the Gaussian action is $\tau$-independent and has a quadratic form in terms of the original fields $(A_\mu, B, \bar{c}, c, \phi,\phi^*)$ as
\begin{multline}
\mc{S}_0^{(2)} \coloneqq \int_k 
 \biggl[
 -\frac{1}{2} A_\mu(k)  \Big( S_0^{AA}(k)\Big)_{\mu\nu} A_\nu(-k)
- ik_\mu A_\mu(k) S_0^{AB}(k) B(-k)
+ \frac{1}{2} B(k) S_0^{BB}(k) B(-k) \\
-  \bar c (k) S_0^{\bar c c}(k) c(-k)
-  \phi^*(k)S_0^{\phi^*\phi}(k) \phi(-k)
\biggr]\,.
\label{eq: Gaussian part with original fields}
\end{multline}
Then, the solution to GF-ERG \cref{e:GF-ERGeq} under the above ansatz yields the explicit forms of the operators $S_0^{(2)}(k)$ in terms of the original fields $(A_\mu, B, \bar{c}, c, \phi)$. 

Next, we define the ``$-1$ variables"~\cite{Miyakawa:2023yob} as
\begin{subequations}
\label{eq: redefined fields}
\begin{align}
    \mc{A}_\mu(k) &\coloneqq e^{k^2} e^{-S_0}\hat{s}^{-1}A_\mu(k)\hat{s}e^{S_0}
    = e^{-k^2} \left(h_{\mu\nu}(k) A_\nu(k)  + ik_\mu h_B(k^2)  B(k)\right)\,,\\[1ex]
    \mc{B}(k) &\coloneqq e^{k^2} e^{-S_0}\hat{s}^{-1}B(k)\hat{s}e^{S_0}
    = e^{-k^2}h_B(k^2) \left(B(k) + e^{2k^2} ik_\mu A_\mu(k) \right)\,,\\[1ex]
    \overline{C}(k) &\coloneqq e^{k^2} e^{-S_0}\hat{s}^{-1}\bar{c}(k)\hat{s}e^{S_0}
    = e^{-k^2} h_S(k^2)\overline{c}\,,\\[1ex]
    C(k) &\coloneqq e^{k^2} e^{-S_0}\hat{s}^{-1}c(k)\hat{s}e^{S_0}
    = e^{-k^2} h_S(k^2)c\,,\\[1ex]
    \Phi(k) &\coloneqq e^{k^2} e^{-S_0}\hat{s}^{-1}\phi(k)\hat{s}e^{S_0}
    =e^{-k^2} h_S(k^2)\phi\,.
\end{align}
\end{subequations}
These variables play an important role as fundamental variables in the perturbative expansion performed later.
Here, we have defined the perturbative propagators $(h_{\mu\nu}(k), h_B(k^2), h_S(k^2))$ of the fields as
\begin{align}
   h_{\mu\nu}(k) & \coloneqq h_S(k^2) T_{\mu\nu}(k) +  \left(\xi + e^{-2k^2} \right) h_B(k^2)L_{\mu\nu}(k)\,,\\
   h_B(k^2) & \coloneqq \frac{1}{k^2+\xi e^{-2k^2}+e^{-4k^2}}\,,\\
    h_S(k^2) & \eqqcolon \frac{1}{k^2+e^{-2k^2}}\,,
\end{align}
where the gauge fixing parameter $\xi$ is an arbitrary real number and the transverse and longitudinal operators $T_{\mu\nu}(k), L_{\mu\nu}(k)$ are defined as,
\begin{align}
\label{eq: projectors}
&T_{\mu\nu}(k)  \coloneqq \delta_{\mu\nu}-\frac{k_\mu k_\nu}{k^2}\,,\quad 
&L_{\mu\nu}(k) \coloneqq \frac{k_\mu k_\nu}{k^2}\,.
\end{align}

Finally, we rewrite \cref{eq: Gaussian part with original fields} in terms of the $-1$ variables \eqref{eq: redefined fields} to obtain
\begin{multline}
\mc{S}_0^{(2)} =  \int_k 
 \biggl[
 -\frac{1}{2} \mc{A}_\mu(k)  \Big( S_0^{\mc{A}\mc{A}}(k)\Big)_{\mu\nu} \mc{A}_\nu(-k)
- ik_\mu \mc{A}_\mu(k) S_0^{\mc{A}\mc{B}}(k) \mc{B}(-k)
+ \frac{1}{2} \mc{B}(k) S_0^{\mc{B}\mc{B}}(k) \mc{B}(-k)\\
-  \overline{C} (k) S_0^{\overline{C} C}(k) C(-k)
-  \Phi^*(k)S_0^{\Phi^*\Phi}(k) \Phi(-k)
\biggr]\,,
\label{eq: Gaussian part with -1 variable}
\end{multline}
where the operators $S_0^{(2)}(k)$ are given by
\begin{subequations}
\label{e:GaussianFPAction}
\begin{align}
 S_0^{\mc{A}\mc{A}}(k)
 &= \vcenter{\hbox{\includegraphics[width=25mm]{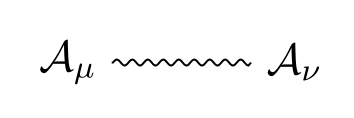}}}
 =  k^2 \qty (1+k^2 e^{2k^2}) T_{\mu\nu}(k) - k^2 e^{2k^2} L_{\mu\nu}(k) \,,
 \label{eq: inverse propagator of A}
 \\
 S_0^{\mc{A}\mc{B}}(k)
 &= \vcenter{\hbox{\includegraphics[width=25mm]{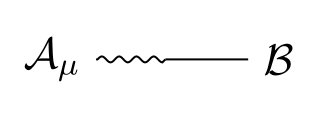}}}
 =  1+\xi e^{2k^2} \,,
 \\
 S_0^{\mc{B}\mc{B}}(k)
 &= \vcenter{\hbox{\includegraphics[width=25mm]{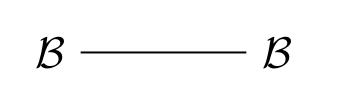}}}
 = \xi + (\xi^2-k^2)e^{2k^2}\,,
\\
S_0^{\Phi^*\Phi}(k) 
 &= \vcenter{\hbox{\includegraphics[width=25mm]{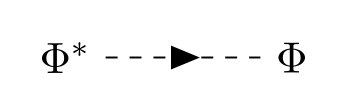}}}
= k^2 \qty(1+k^2e^{2k^2}),
\\
S_0^{\overline CC}(k)
 &= \vcenter{\hbox{\includegraphics[width=25mm]{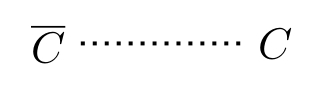}}}
 = k^2 \qty(1+k^2e^{2k^2})\,.
\end{align}
\end{subequations}

\begin{figure}
\centering
\includegraphics[width=0.5\linewidth,pagebox=cropbox,clip]{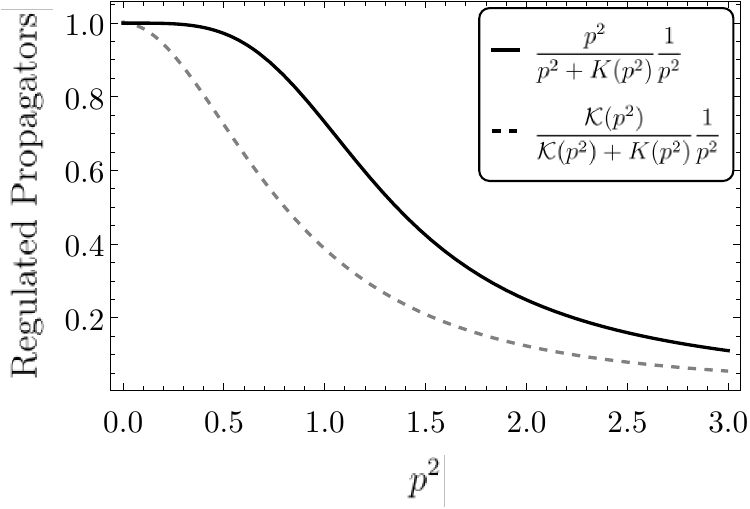}
\caption{Comparison of the regulated propagators between $\mathcal K(p^2)=p^2$ (solid line) and \cref{eq: cutoff function mathcal K} with $K(p^2)=e^{-p^2}$ (dashed line).}
\label{fig:ComparisonPropagators}
\end{figure}

Here, we discuss the consistency of the choice \eqref{e:ChoiceOfCalK} for $\mc{K}(p^2)$ as an ultraviolet (UV) cutoff function.
It should be noted that $h_S(k^2)$ plays the role of the loop propagator of the matter field in the following perturbative analysis, whose momentum dependence is depicted in \Cref{fig:ComparisonPropagators}.
There is no large difference in the qualitative behaviors of the regulated propagator between the choice \eqref{e:ChoiceOfCalK} and \cref{eq: cutoff function mathcal K} with the choice \eqref{eq: cutoff function K}.
Therefore, the choice \eqref{e:ChoiceOfCalK} can be regarded reliable for playing a role as a UV cutoff function, at least at the level of perturbative analysis.

Also, we discuss here gauge-fixing of the action.
Because this Gaussian action is a fixed point of the GF-ERG equation with $e_\tau = 0$, we can take the continuum limit ($\Lambda_0 \to \infty$).
After the limit, we get
\begin{multline}
 S_0\to  S_G \coloneqq \int_k 
 \biggl[
 - \frac{1}{2} A_\mu(k) \qty( k^2 T_{\mu\nu}(k) + \frac{k^2}{1+\xi}L_{\mu\nu}(k))A_\nu(k) 
 - \frac{1}{1+\xi} B(-k) i k_\mu A_\mu(k)
 + \frac{\xi}{2(1+\xi)} B(-k)B(k)
 \\ 
 - \bar{c}(-k) k^2 c(k)  - \phi^*(k)k^2\phi(k)
 \biggr],
\end{multline}
which is expressed in the coordinate space as 
\begin{multline}
 S_G = 
 \int_x 
 \biggl[
 - \frac{1}{4} F_{\mu\nu}(x)F_{\mu\nu}(x)
 - \frac{1}{2(1+\xi)} (\del_\mu A_\mu(x))^2
 - \frac{1}{1+\xi} B(x) \del_\mu A_\mu(x)
 + \frac{\xi}{2(1+\xi)} B(x)B(x)
 \\ 
 - \del_\mu\bar{c}(x) \del_\mu c(x)  - \del_\mu \phi^* (x) \del_\mu\phi(x)
 \biggr].
\end{multline}
If we integrate out the NL field $B$, we get
\begin{align}
 S_G \to
 \int_x 
 \biggl[
 - \frac{1}{4} F_{\mu\nu}(x)F_{\mu\nu}(x)
 - \frac{1}{2 \xi} (\del_\mu A_\mu(x))^2
 - \del_\mu\bar{c}(x) \del_\mu c(x)  - \del_\mu \phi^* (x) \del_\mu\phi(x)
 \biggr],
\end{align}
which agrees with the continuum sQED action in the $R_\xi$ gauge.
Therefore, we can consider \Cref{eq: Gaussian part with original fields} or \Cref{eq: Gaussian part with -1 variable} as the free sQED action in the $R_\xi$ gauge modified by introducing the UV cutoff.

\subsection{Ward-Takahashi identity}
\label{sec: Ward-Takahashi identity}
We discuss the Ward-Takahashi identities here before deriving the flow equations for the effective vertices $S_\tau^{(n)}$ from the GF-ERG equation \eqref{e:GF-ERGeq} within the perturbative expansion.
In the conventional Wilson-Polchinski equation, there are two origins of the breaking of BRST symmetry: One is the diffusion equation for the fields and the other originates from the scrambler operator $\hat s$.
The former can be solved by modifying the diffusion equations to \Cref{eq: GF diffusion equations}, as was discussed in \Cref{sec: Definition of Wilsonian effective action}.
Indeed, the latter makes the GF-ERG equation \eqref{e:GF-ERGeq} non-invariant under the original BRST transformation \eqref{eq: BRST transformation}, which is generated by
\begin{align}
\hat{\mathcal G}_B \coloneqq \int_p
\biggl[ ip_\mu c(p)\fdv{A_\mu(p)} + B(p) \fdv{\bar c(p)}
+ ie_\tau c(p) \phi(p) \fdv{\phi(p)} - ie_\tau c(p) \phi^*(p) \fdv{\phi^*(p)}
\biggr]\,.
\end{align}
This is because the scrambler operator is not commutative with this BRST generator, i.e. $[\hat s, \hat{\mathcal G}_B ]\neq0$.
Instead, following \cite{Miyakawa:2021wus,Miyakawa:2021hcx}, we can show that the GF-ERG equation \eqref{e:GF-ERGeq} is invariant under the modified BRST transformation, whose generator reads
\begin{align}
\tilde{\hat{\mathcal G}}_B & \coloneqq \hat{s} \hat{\mathcal G}_B  \hat{s}^{-1}
 =\int_p \biggl[ \del_\mu \widehat c(p)\fdv{A_\mu(p)} 
+ \widehat B(p) \fdv{\bar c(p)} 
 + ie_\tau \widehat c(p) \phi(p) \fdv{\phi(p)} 
- ie_\tau \widehat c(p) \phi^*(p) \fdv{\phi^*(p)}
\biggr]\,,
\label{eq: modified BRST generator}
\end{align}
where the fields with hat are defined as
\begin{align}
&\widehat {c}(p) =c(p)+\fdv{\bar c(p)}\,,&
&\widehat {\bar c}(x) =\bar c(p)- \fdv{c(p)}\,,&
&\widehat {B}(p) = B(p) - \fdv{B(p)}\,.
\label{eq: hat fields for unphysical fields}
\end{align}
The invariance of the system under the (modified) BRST transformation \eqref{eq: modified BRST generator} leads to the modified Ward-Takahashi (WT) identities for the effective vertices $S^{(n)}_\tau$.
The explicit form of the modified WT identities is lengthy, so we give them in \Cref{app: Modified BRST invariance}. 
In the next section, we employ the perturbative analysis in the polynomial of $e_\tau$ on the effective vertices. In such a case, the Ward-Takahashi identity for the Wilsonian effective action is reduced to 
\begin{align}
\label{e:rWTid}
e^{k^2} k_\mu \frac{\delta (S_\tau-\mc{S}^{(2)}_0)}{\delta\mathcal{A}_\mu(k)} + e_\tau \int_p 
\qty (
    \phi(p) \fdv{S_\tau}{\phi(p+k)} - \phi^*(p) \fdv{S_\tau}{\phi^*(p+k)}
) = 0,
\end{align}
where $\mc{S}_0^{(2)}$ is the Gaussian part (tree-level part of the Wilsonian effective action $S_\tau$) defined in \Cref{eq: Gaussian part with -1 variable}.
The detailed discussion is given in \Cref{app: Modified BRST invariance}.
Note that in \cref{e:rWTid}, the first term is given by the functional derivative with respect to the $-1$ variable ($\mc{A}_\mu$), while the second and third terms are given with the original fields ($\phi$ and $\phi^*$).

\section{Perturbative analysis of GF-ERG equation}
\label{sec: Perturbative analysis of GF-ERGeq}
In this section, we intend to solve the GF-ERG equation \cref{e:GF-ERGeq} using perturbation theory.
To this end, we suppose here that the $n$-point correlation functions $S^{(n)}_\tau$ are functions of solely the renormalized charged coupling $e_\tau$, and thus $S^{(n)}_\tau$ are expanded into the polynomial of $e_\tau$ as
\begin{align}
S^{(n)}_\tau &= \sum_{l=0}^\infty (e_\tau)^l S^{(n)}_l
= S^{(n)}_0 + e_\tau S^{(n)}_1 + e_\tau^2 S^{(n)}_2 +\cdots \,.
\label{eq: perturbative expansion of S}
\end{align}
Furthermore, assuming that the explicit dependence of $\tau$ on $S^{(n)}_\tau$ is given only in the coupling $e_\tau$, the left-hand side of \cref{e:GF-ERGeq} can be rewritten in terms of the change of $e_\tau$ such that
\begin{align}
\partial_\tau S_\tau &= \partial_\tau e_\tau \frac{\partial S_\tau}{\partial e_\tau}
 = \left[\frac{4-D}{2}-\gamma_A(\tau) \right] e_\tau \frac{\partial S_\tau}{\partial e_\tau}\,,
\end{align}
where we have used the renormalization group equation for $e_\tau$:
\begin{align}
\partial_\tau e_\tau=\left[\frac{4-D}{2}-\gamma_A(\tau) \right] e_\tau\,,
\end{align}
which follows from the definition (\Cref{eq: definition of renormalized charge}) of the renormalized charge $e_\tau$.

Then, we evaluate the GF-ERG equation for each vertex order by order of $e_\tau$.
In this work, we consider the quantum corrections up to the second order of $e_\tau$.
We focus on the two-point correlation function for the gauge fields and compute $O(e_\tau^2)$ to it.
On the other hand, we just consider the Gaussian part for the other fields: that is, the zeroth order of $e_\tau$ is considered\footnote{
This assumption is justified because the ghosts and Nakanishi-Lautrup fields decouple from the gauge and matter fields in the linear covariant gauge such as the $R_\xi$ gauge.
}.
Furthermore, we do not consider the $(\phi^*\phi)^2$-vertex and two-point function of $\phi\phi^*$ on the order of $e_\tau^{2}$.
More specifically, we assume the following ansatz for Wilsonian effective action $S_\tau$:
\begin{multline}
S_\tau =  {\mathcal S}_0^{(2)} +\int_{p,k}  \Phi^*(p+k)\Phi(p) \left( e_\tau S^{\Phi^*\Phi \mc{A}}_1(p,k) \right)_\mu \mc{A}_\mu (k) 
\\
+ \int_{p,k,l} \Phi^*(p+k+l) \Phi(p) \mc{A}_\mu(k)  \left( e_\tau^2S_2^{\Phi^*\Phi\mc{A}\mc{A}}(p,k,l) \right)_{\mu\nu}\mc{A}_\nu(l)
+\int_k 
 \frac{1}{2} \mc{A}_\mu(k)   \Big(e_\tau^2S_2^{\mc{A}\mc{A}}(k) \Big)_{\mu\nu} \mc{A}_\nu(-k)
\,,
\label{eq: perturbative expansion of effective action}
\end{multline}
where ${\mathcal S}_0^{(2)}$ is the Gaussian part given in \Cref{eq: Gaussian part with -1 variable}.
This ansatz is minimal for computing the quantum corrections to the two-point function of the gauge fields at the second order because the other terms do not contribute to it at this order. 
In the following, we construct $S_l^{(n)}$ order by order in the Wilsonian effective action~\eqref{eq: perturbative expansion of effective action}.

\subsection{First order \texorpdfstring{$O(e_\tau)$}{}}
\label{sec: First order}
We next deal with the solution of the GF-ERG equation at the first order of $e_\tau$, i.e. $S^{\Phi^*\Phi \mc{A}}_1(p,k)$ in \Cref{eq: perturbative expansion of effective action}:
\begin{align}
\label{e:AnsatzS1}
\mc{S}_1^{\Phi^*\Phi \mc{A}}\coloneqq\int_{p,k} {\Phi}^*(p+k)  {\Phi}(p)\left( e_\tau S^{\Phi^*\Phi \mc{A}}_1(p,k) \right)_\mu {\mc{A}}_\mu (k)\,.
\end{align}
In terms of $\Phi$, $\Phi^*$ and $\mc{A}_\mu$ as given in \Cref{eq: redefined fields}, the term $\mc{S}_1^{\Phi^*\Phi \mc{A}}$ in \Cref{eq: perturbative expansion of S} obeys
\begin{multline}
-\frac{D-4}{2} \mc{S}^{\Phi^*\Phi \mc{A}}_1
-\int_k 
\Biggl[
     \qty(k\cdot \del_k + \frac{D+2}{2}) \mc{A}_\mu(k) \cdot \fdv{\mc{S}^{\Phi^*\Phi \mc{A}}_1}{\mc{A}_\mu(k)}
    +
    \Biggl(
        \qty(k\cdot \del_k + \frac{D+2}{2}) \Phi(k) \cdot \fdv{\mc{S}^{\Phi^*\Phi \mc{A}}_1}{\Phi(k)}
        + \text{(c.c.)}
    \Biggr)
   \Biggr]
   \\
=4\int_{p,k}\,
	\qty( e^{p^2-(p+k)^2-k^2} p^2 (p+k)_\mu+ e^{(p+k)^2-p^2-k^2} (p+k)^2 p_\mu)\Phi^*(p+k) \mc{A}_\mu (k) \Phi(p)\,.
\label{e:1stOrder}
\end{multline}
The anomalous dimensions of $\mathcal A_\mu$ and $\Phi$, denoted respectively by $\gamma_A$ and $\gamma_{\phi}$, are assumed to be of order of $e_\tau^2$, so that those are neglected here in $\mc{S}^{\Phi^*\Phi \mc{A}}_1$.
Note that the first two terms on the right-hand side of \Cref{e:1stOrder} exist in the conventional Wilson-Polchinski equation, while the last term arises from modification of the gradient flow equations.

\begin{figure}
\centering
\includegraphics[width=0.5\linewidth,pagebox=cropbox,clip]{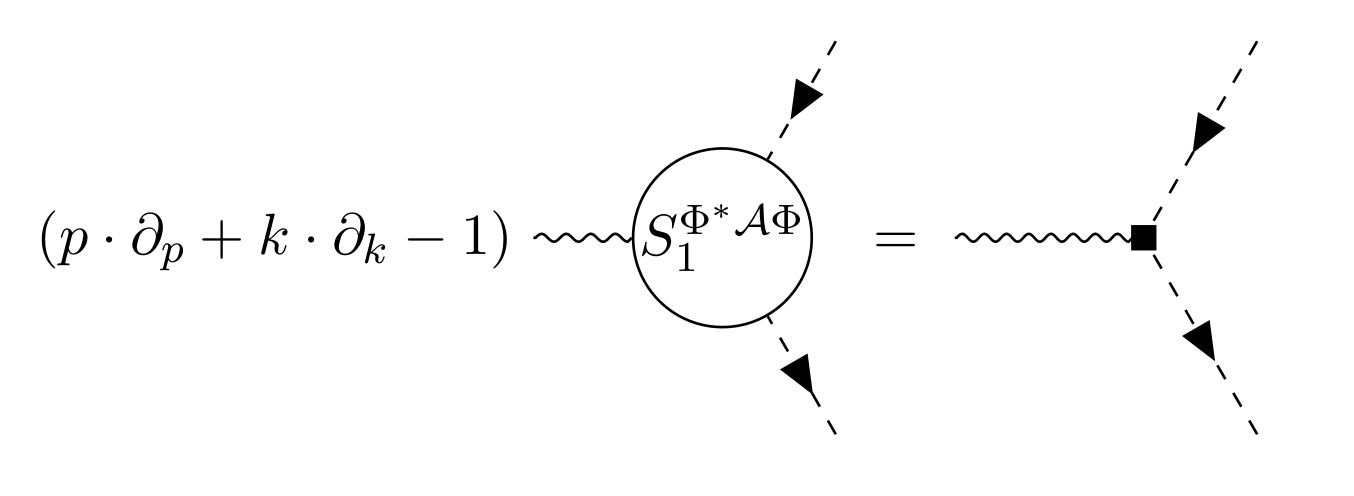}
\caption{Diagrammatic representation of the GF-ERG equation for $(S_{1}^{\Phi^*\Phi\mc{A}})_{\mu}$ given in \cref{e:CondVmu}.
 The wiggly line represents the propagator of the gauge field $A_\mu$, while the complex scalar fields are shown by the dashed line with an arrow. The first two terms on the right-hand side are involved in the conventional Wilson-Polchinski equation, while the others originate from modification of the diffusion equation~\eqref{eq: GF diffusion equations}. The black circle and square dots represent the additional vertices that arise from interactions in the modified diffusion equation~\eqref{eq: GF diffusion equations} or, equivalently, the last term in the GF-ERG \eqref{e:GF-ERGeq}.
}
\label{fig:three-point function}
\end{figure}

Substituting the ansatz \Cref{e:AnsatzS1} for $\mc{S}^{\Phi^*\Phi \mc{A}}_1$ into \Cref{e:1stOrder}, we obtain the equation for the vertex $(S^{\Phi^*\Phi \mc{A}}_1(p,k))_\mu$ such that
\begin{align}
\qty(p\cdot\del_p+k\cdot\del_k-1) \left(S^{\Phi^*\Phi \mc{A}}_1(p,k) \right)_\mu =
   4 \biggl(e^{p^2-(p+k)^2-k^2}p^2 (p+k)_\mu 
+ e^{(p+k)^2-p^2-k^2}(p+k)^2 p_\mu \biggr)\,.
\label{e:CondVmu}
\end{align}
The left- and right- hand sides of \Cref{e:CondVmu} correspond to those of \Cref{e:1stOrder}, respectively.
We show the diagrammatic expression for \Cref{e:CondVmu} in \Cref{fig:three-point function}. 
Using the formula in Appdix A in \cite{Miyakawa:2021wus}, the general solution to \cref{e:CondVmu} is found to be
\begin{align}
\label{e:VmuWithAB}
\left(S^{\Phi^*\Phi \mc{A}}_1(p,k) \right)_\mu
 = ap_\mu + b k_\mu 
 +2 \biggl(F(p^2-(p+k)^2-k^2)p^2 (p+k)_\mu 
+ F((p+k)^2-p^2-k^2)(p+k)^2p_\mu\biggr)\,,
\end{align}
where $a$ and $b$ are arbitrary constants, and we have introduced a function
\begin{align}
\label{eq: function F}
F(x)\coloneqq \frac{e^x-1}{x}\,.
\end{align}
The Ward-Takahashi identity constrains $a$ and $b$ to be $2$ and $1$, respectively, as will be discussed in \Cref{app: Modified BRST invariance}.
Therefore, the three-point vertex $\left(S^{\Phi^*\Phi \mc{A}}_1(p,k) \right)_\mu$ is explicitly obtained such that 
\begin{align}
\label{e:Vmu}
\left(S^{\Phi^*\Phi \mc{A}}_1(p,k) \right)_\mu
=  (2p + k)_\mu + 2F(p^2-(p+k)^2-k^2)p^2 (p+k)_\mu 
 + 2F((p+k)^2-p^2-k^2)(p+k)^2p_\mu\,.
\end{align}

\subsection{Second order \texorpdfstring{$O(e_\tau^2)$}{}}
We construct these terms by means of GF-ERG \cref{e:GF-ERGeq} here in the same manner in \Cref{sec: First order}.
The effective action \eqref{eq: perturbative expansion of effective action} at second order of $e_\tau^2$ is assumed to contain the $\Phi^* \Phi\mc{A}\mc{A}$ and $\mc{A}\mc{A}$ terms:
\begin{align}
\mc{S}_2^{\Phi^*\mc{A}\mc{A}\Phi} &\coloneqq  \int_{p,k,l} \Phi^*(p+k+l) \Phi(p) \mc{A}_\mu(k) \left( e_\tau^2S_2^{\Phi^*\Phi\mc{A}\mc{A}}(p,k,l) \right)_{\mu\nu}  \mc{A}_\nu(l)\,,
\label{eq: AAphiphi term}
\\
\mc{S}_2^{\mc{A}\mc{A}} &\coloneqq  \frac{1}{2} \int_k \mc{A}_\mu(k)  \Big( e_\tau^2 S_2^{\mc{A}\mc{A}}(k) \Big)_{\mu\nu} \mc{A}_\nu(-k)\,.
\end{align}

\subsubsection{Four-point correlation function}
\begin{figure}
\centering
\includegraphics[width=1.0\linewidth,pagebox=cropbox,clip]{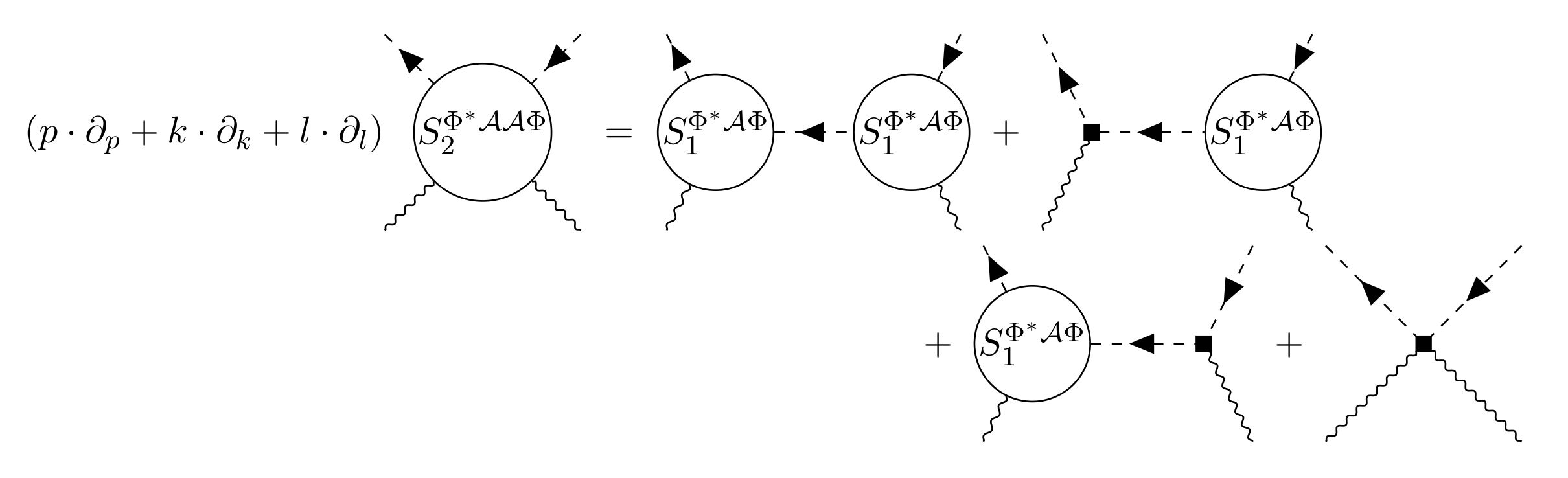}
\caption{Diagrammatic representation of the GF-ERG equation for $(S_{2}^{\Phi^*\mc{A}\mc{A}\Phi})_{\mu\nu}$ given in \Cref{eq: ERGE for four-point vertex}.
The Feynman rules for the diagrams are the same as \Cref{fig:three-point function}. The conventional Wilson-Polchinski equation has the first two terms on the right-hand side, while the modification of the diffusion equation~\eqref{eq: GF diffusion equations} entails the others.
}
\label{fig:four-point function}
\end{figure}

From the GF-ERG equation, the four-point function $\mc{S}_2^{\Phi^* \Phi\mc{A}\mc{A}}$ obeys
\begin{multline}
-(D-4)\mc{S}_{2}^{\Phi^*\Phi \mc{A}\mc{A}}
- \int_p
\left[
    \qty(p\cdot \p_p+\frac{D+2}{2})\mc{A}_\mu(p) \cdot \fdv{\mc{S}_{2}^{\Phi^* \Phi\mc{A}\mc{A}}}{\mc{A}_\mu(p)}
    + \biggl(
        \qty(p\cdot \p_p+\frac{D+2}{2})\Phi(p)\cdot \fdv{\mc{S}_{2}^{\Phi^*\Phi\mc{A}\mc{A}}}{\Phi(p)} + \text{(c.c.)}
    \biggr)
\right] \\
= \int_p 2(2p^2+1)\fdv{\mc{S}_1^{\Phi^*\Phi \mc{A}}}{\phi^*(p)}\fdv{\mc{S}_1^{\Phi^*\Phi \mc{A}}}{\phi(p)}
\\
+ \int_x 
\biggl[
    4i\fdv{\mc{S}_1^{\Phi^*\Phi \mc{A}}}{\phi^*(x)}\qty(A_\mu(x)+\fdv{\mc{S}_0^{(2)} }{A_\mu(x)})\del_\mu \qty(\phi^*(x) +\fdv{\mc{S}_0^{(2)}}{\phi(x)})
    + 4i \fdv{\mc{S}_0^{(2)}}{\phi^*(x)}\qty(A_\mu(x) +\fdv{\mc{S}_0^{(2)}}{A_\mu(x)})\del_\mu \fdv{\mc{S}_1^{\Phi^*\Phi \mc{A}}}{\phi(x)}
    \\
    -2 \fdv{\mc{S}_0^{(2)}}{\phi^*(x)}\qty(A_\mu(x) +\fdv{\mc{S}_0^{(2)}}{A_\mu(x)})\qty(A_\mu(x) +\fdv{\mc{S}_0^{(2)}}{A_\mu(x)})\phi(x)
    +(\text{c.c.})
\biggr]\,.
\label{e:ppAA}
\end{multline}
Note here that functional derivatives on the left-hand side \Cref{e:ppAA} are given by the $-1$ variables, and those on the right-hand side are given by the original fields.
Those functional derivatives can be rewritten in each other using the chain rule, e.g.,
$
{\delta}/{\delta \phi}={\delta \Phi}/{\delta \phi} \cdot {\delta}/{\delta \Phi}
$,
where the relation between the original fields and $-1$ variables are given in \Cref{eq: redefined fields}.
Substituting the ansatz \eqref{eq: AAphiphi term} into \Cref{e:ppAA} yields
\begin{multline}    
\label{eq: ERGE for four-point vertex}
\qty(p\cdot \p_p + k\cdot \p_k + l\cdot \p_l)\left(S_2^{\Phi^*\Phi\mc{A}\mc{A}}(p,k,l)\right)_{\mu\nu} 
 = 
2(2(p+k)^2+1)\left(S^{\Phi^*\Phi \mc{A}}_1(p,k) \right)_\mu e^{-2(p+k)^2}(h_S(p+k))^2 \left(S^{\Phi^*\Phi \mc{A}}_1(p+k,l) \right)_\nu
\\
- 4\Biggr(
	(p+k+l)_\nu\left(S^{\Phi^*\Phi \mc{A}}_1(p,k) \right)_\mu e^{-(p+k)^2-(p+k+l)^2-l^2} h_S(p+k)
 +
	p_\mu \left(S^{\Phi^*\Phi \mc{A}}_1(p+k,l)  \right)_\nu e^{-(p+k)^2 - p^2 - k^2} h_S(p+k)
	\Biggl)
\\
+ 4\Biggr(
	 p^2 (p+k)_\mu e^{p^2-(p+k)^2-k^2} h_S(p+k) \left(S^{\Phi^*\Phi \mc{A}}_1(p+k,l)  \right)_\nu
 	+
 	\left(S^{\Phi^*\Phi \mc{A}}_1(p,l)  \right)_\mu h_S(p+k) (p+k+l)^2 (p+k)_\nu e^{(p+k+l)^2-(p+k)^2 -l^2} 
	\Biggl) 
\\
 - 2 \delta_{\mu\nu} e^{-k^2-l^2}\qty[e^{(p+k+l)^2-p^2} (p+k+l)^2 + e^{p^2-(p+k+l)^2} p^2]\,,
\end{multline}
where $\left( S^{\Phi^*\Phi \mc{A}}_1(p,l)  \right)_\mu$ is given in \Cref{e:Vmu}.
We have used the explicit forms of the propagators \eqref{e:GaussianFPAction}. 
Its diagrammatic representation is depicted in \Cref{fig:four-point function}. 
The general solution to this equation is found to be
\begin{multline}
\left( S_2^{\Phi^*\Phi\mc{A}\mc{A}}(p,k,l) \right)_{\mu\nu} 
= c \delta_{\mu\nu} + \left(S^{\Phi^*\Phi \mc{A}}_1(p,k) \right)_\mu h_S(p+k) \left(S^{\Phi^*\Phi \mc{A}}_1(p+k,l) \right)_\nu
\\
- \delta_{\mu\nu} \biggl( (p+k+l)^2 F((p+k+l)^2-p^2-k^2-l^2)
 + p^2 F(p^2 -(p+k+l)^2 -k^2-l^2 )
\biggr) 
-4X_{\mu\nu}(p,k,l)\,,
\label{e: SphiphiAA}
\end{multline}
where $c$ is an arbitrary real number and $F(x)$ is defined \cref{eq: function F}.
$X_{\mu\nu}(p,k,l)$ is a particular solution to the following equation:
\begin{align}
\label{e:DiffEqXmunu}
&\qty(p\cdot \p_p + k\cdot \p_k + l\cdot \p_l) X_{\mu\nu}(p,k,l)\nn
&\qquad
= 
	(p+k+l)_\nu \left(S^{\Phi^*\Phi \mc{A}}_1(p,k) \right)_\mu e^{(p+k)^2-(p+k+l)^2-l^2}
	+
	p_\mu \left(S^{\Phi^*\Phi \mc{A}}_1(p+k,l) \right)_\nu e^{(p+k)^2 - p^2 - k^2}\,,
\end{align}
whose concrete form can be obtained exactly, but is very lengthy, so is given in \Cref{e:Xmunu} in \Cref{a:Xmunu}.
The Ward-Takahashi identity \eqref{e:rWTid} determines the arbitrary constant $c$ in \Cref{e: SphiphiAA} to be $-1$, which is discussed in \Cref{app: Modified BRST invariance}.
Then, $\left( S_2^{\Phi^*\Phi\mc{A}\mc{A}}(p,k,l) \right)_{\mu\nu}$ is given by
\begin{multline}
\label{e:Vmunu}
\left( S_2^{\Phi^*\Phi\mc{A}\mc{A}}(p,k,l) \right)_{\mu\nu}  = - \delta_{\mu\nu} +  \left(S^{\Phi^*\Phi \mc{A}}_1(p,k) \right)_\mu h_S(p+k)  \left(S^{\Phi^*\Phi \mc{A}}_1(p+k,l) \right)_\nu
\\
- \delta_{\mu\nu} \biggl( (p+k+l)^2 F((p+k+l)^2-p^2-k^2-l^2) 
 + p^2 F(p^2 -(p+k+l)^2 -k^2-l^2 )\biggr) 
-4X_{\mu\nu}(p,k,l)\,.
\end{multline}

\subsubsection{Two-point function of gauge fields}
\begin{figure}
\centering
\includegraphics[width=1.0\linewidth,pagebox=cropbox,clip]{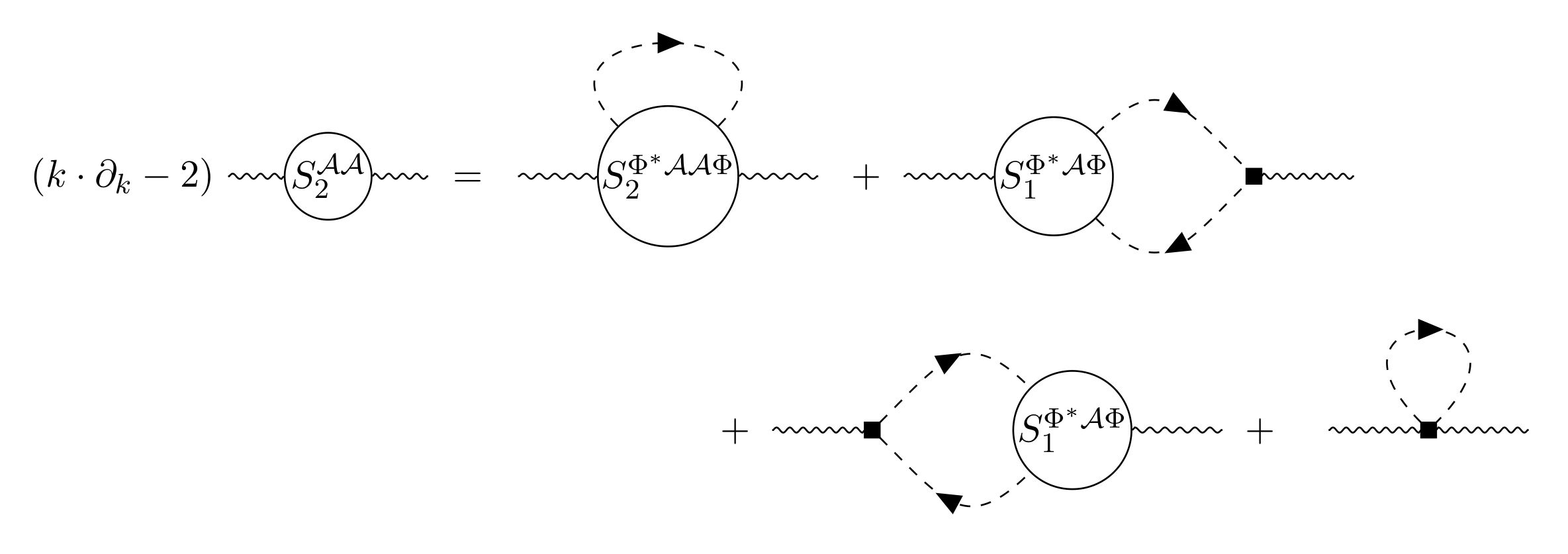}
\caption{The diagrammatic representation of the GF-ERG equation for $\left( S_2^{\mc{A}\mc{A}}(k) \right)_{\mu\nu}$ in $D=4$.
The Feynman rules for the diagrams are the same as \Cref{fig:three-point function}.
}
\label{fig:two-point function of e2}
\end{figure}

Using the three-point vertex \eqref{e:Vmu} and the four-point vertex \eqref{e:Vmunu}, let us finally analyze the two-point correlation function of the gauge field
\begin{align}
\mc{S}_2^{\mc{A}\mc{A}} = \frac{1}{2} \int_k \mc{A}_\mu(k) \left( e_\tau^2 S_2^{\mc{A}\mc{A}}(k)\right)_{\mu\nu} \mc{A}_\nu(-k)\,.
\label{eq: second order of AA term}
\end{align}
Here, we assume $\gamma_A = \gamma_1 e_\tau^2 + \order{e_\tau^4}$ and then the GF-ERG equation for $\mc{S}_2^{\mc{A}\mc{A}}$ is given by
\begin{align}
&-(D-4)\mc{S}_2^{\mc{A}\mc{A}} -\int_k \qty[
\qty(k\cdot \p_k+\frac{D+2}{2})\mc{A}_\mu(k) \cdot \fdv{\mc{S}_2^{\mc{A}\mc{A}}}{\mc{A}_\mu(k)} - \gamma_1 \qty(A_\mu(k) +\fdv{\mc{S}_0^{(2)}}{A_\mu(-k)}) \fdv{\mc{S}_0^{(2)}}{A_\mu(k)}
]\nn
&\quad
= \int_p 2(2p^2+1)\sfdv{\mc{S}_{2}^{\Phi^*\mc{A}\mc{A}\Phi}}{\phi^*(p)}{\phi(p)}
 \nn
&\qquad + \int_x
\Biggl[
    4i\qty(A_\mu(x) +\fdv{\mc{S}_0^{(2)}}{A_\mu(x)}) \fdv{\phi^*(x)}\del_\mu \fdv{\mc{S}_1^{\Phi^*\Phi \mc{A}}}{\phi(x)}
    +2 \qty(A_\mu(x) +\fdv{\mc{S}_0^{(2)}}{A_\mu(x)})\qty(A_\mu(x) +\fdv{\mc{S}_0^{(2)}}{A_\mu(x)}) \fdv{\Phi(x)}{\phi(x)}
    +(\text{c.c.})
\Biggr].
 \label{e:AA2}
\end{align}
Inserting \Cref{eq: second order of AA term} into \Cref{e:AA2}, we obtain
\begin{align}
&\qty[k\cdot \p_k - 2 + (4-D)]\left( S_2^{\mc{A}\mc{A}}(k)\right)_{\mu\nu} - 2 \gamma_1 k^2 T_{\mu\nu}(k) \nn
&
= \int_p 2(2p^2+1) e^{-2p^2}h_S^2(p) \qty( \left(S_2^{\Phi^*\Phi\mc{A}\mc{A}}(p,k,-k) \right)_{\mu\nu} +  \left(S_2^{\Phi^*\Phi\mc{A}\mc{A}}(p,-k,k) \right)_{\nu\mu})
\nn
&\quad 
- 4 e^{-k^2}\int_p e^{-(p+k)^2-p^2} h_S(p+k)h_S(p)
\left[ \left(S^{\Phi^*\Phi \mc{A}}_1(p,k) \right)_\mu(2p+k)_\nu + \left(S^{\Phi^*\Phi \mc{A}}_1(p,k) \right)_\nu (2p+k)_\mu \right]\nn
&\qquad
+ 8 e^{-2k^2} \delta_{\mu\nu} \int_p e^{-2p^2}h_S(p)\,,
\label{e:eqVAsym}
\end{align}
whose diagrammatic representation is shown in \Cref{fig:two-point function of e2}.
Here, we have symmetrized momenta and Lorentz indices on the right-hand side under $k \leftrightarrow -k$ and $\mu \leftrightarrow \nu$ because $\left( S_2^{\mc{A}\mc{A}}(k)\right)_{\mu\nu}$ is invariant under those replacements. We have also changed the integration variable $p \to -p$ and used the fact that $h_S(-q)=h_S(q)$, to rewrite the second term in the third line of the above equation.

Finding the solution to \Cref{e:eqVAsym} for $( S_2^{\mc{A}\mc{A}}(k))_{\mu\nu}$ is very complicated. Instead, we analyze here the solution in the polynomial expansion in the external momentum $k$. After tedious calculations, we find that the polynomial expansion in $k$ on the right-hand side of \Cref{e:eqVAsym} in $D=4$ is given by  
\begin{align}
\label{e:ExpansionDiffVmunu}
\text{R.H.S. of \Cref{e:eqVAsym}} = \left( S_2^{\mc{A}\mc{A}}(0)\right)_{\mu\nu}  -\frac{2}{48\pi^2} k^2 T_{\mu\nu}(k) + \order{k^4}\,.
\end{align}
Comparing the $O(k^2)$ terms between the two-hand sides of \Cref{e:eqVAsym}, we can read the anomalous dimension of the gauge field as
\begin{align}
\gamma_1 = \frac{1}{48\pi^2}\,,
\end{align}
which leads to the beta function of $e_\tau^2$ at one-loop order,
\begin{align}
\del_\tau e_\tau = - \gamma_1e^3_\tau = - \frac{1}{48\pi^2}e^3_\tau \,.
\end{align}
This result agrees with those of the ordinary perturbative computation in $D=4$.
Note that due to $\del_\tau = - \Lambda \del_\Lambda$, the sign is opposite to the standard expression of the beta function in sQED.
It should be emphasized here that the renormalization factor of the electric charge (beta function) is related to the wavefunction renormalization factor of the photon field (anomalous dimension) through the Ward-Takahashi identity.
Moreover, it is well-established that the one- and two-loop coefficients of the beta function are independent of the RG scheme.
If a gauge-invariant exact renormalization group (ERG) formulation adheres to the ordinary Ward-Takahashi identity, it must accurately reproduce the photon anomalous dimension up to the two-loop level.
Therefore, our result that the one-loop anomalous dimension is correctly reproduced acts as a necessary and significant consistency check for GF-ERG as a gauge-invariant ERG framework.

Let us next investigate the loop correction to the mass term of the photon involved in \cref{e:ExpansionDiffVmunu} in a general spacetime dimension $D$.
To this end, setting $k=0$, we find that \Cref{e:eqVAsym} is reduced to
\begin{align}
\label{eq: SAA2}
\frac{2-D}{2}\left( S_2^{\mc{A}\mc{A}}(0)\right)_{\mu\nu} 
& = \int_p 2(2p^2+1) e^{-2p^2}h_S^2(p)  \left(S_2^{\Phi^*\Phi\mc{A}\mc{A}}(p,0,0) \right)_{\mu\nu}\nn
 &\quad - 4 \int_p e^{-2p^2} \qty[h_S(p) \left(S^{\Phi^*\Phi \mc{A}}_1(p,0) \right)_\nu h_S(p)](2p_\mu)
   + 4\delta_{\mu\nu} \int_p e^{-2p^2}h_S(p)\,.
\end{align}
Using the explicit forms of $(S^{\Phi^*\Phi \mc{A}}_1(p,k))_\mu$ and $( S_2^{\Phi^*\Phi\mc{A}\mc{A}}(p,k,l))_{\mu\nu}$ given in \Cref{e:Vmu} and \Cref{e:Vmunu}, this equation becomes
\begin{align}
\frac{2-D}{2} \left( S_2^{\mc{A}\mc{A}}(0)\right)_{\mu\nu}  
 = - \frac{\Gamma\qty(\frac{D}{2})}{(4\pi)^{D/2}} \frac{\delta_{\mu\nu}}{D} \eval{\qty[\frac{4 \qty(4 (p^2)^2+2 p^2+1-2e^{-2 p^2}) (p^2)^{D/2}e^{-2 p^2}}{(p^2+e^{-2 p^2})^2}]}_{p=0}^{p=\infty}
 = 0\,.
\label{e:NoMassTerm}
\end{align}
Thanks to the Lorentz covariance, $\left( S_2^{\mc{A}\mc{A}}(0)\right)_{\mu\nu} $ must be proportional to $\delta_{\mu\nu}$, and thus its coefficient gives the mass term: $\left( S_2^{\mc{A}\mc{A}}(0)\right)_{\mu\nu} \coloneqq - m^2_A \delta_{\mu\nu}$.
Therefore, \Cref{e:NoMassTerm} means that the mass term is not generated at the 1-loop level. 

This result is in contrast to the case of the conventional Wilson-Polchinski equation where the counterpart of \Cref{e:ExpansionDiffVmunu} reads
\begin{align}
    - m_A^2\delta_{\mu\nu} - \frac{2}{48\pi^2} k^2 T_{\mu\nu}(k) + \order{k^4}\,,
\end{align}
where $m_A^2$ is expressed as the non-vanishing integral given by 
\begin{align}
m_A^2 =  \int_p 2(2p^2+1)e^{-2p^2} h_S^2(p) (D + e^{-2p^2}h_S(p)) > 0.
\end{align}
This expression can be obtained by neglecting the contribution from the extra vertices, which are depicted as square dots in the Feynman diagrams and originate from the non-linear terms in the gradient flow equation in \Cref{eq: GF diffusion equations}.
Comparing both cases, we see that the transversality of the photon two-point function and the anomalous dimension of the photon are consistent with perturbation theory even in the Wilson-Polchinski case, and GF-ERG holds these properties.
This is because the $O(k^2)$-term receives logarithmic corrections $\sim \log k$. 
The striking point is that by changing the flow equation from the simple diffusion one to the gradient flow, the quantum correction to the mass term of the photon two-point function is canceled.
In this sense, the GF-ERG formulation realizes a gauge-invariant renormalization group flow in contrast to the conventional Wilson-Polchinski equation.

\section{Conclusions and Prospects}
\label{sec: conclusions}
In this paper, we have studied scalar Quantum Electrodynamics (sQED) utilizing the Gradient Flow Exact Renormalization Group (GF-ERG) formulation. Its central idea is to focus on diffusion equations (or gradient flow equations) for fields included in the Exact Renormalization Group equation.
The conventional Wilson-Polchinski equation contains simple diffusion equations like $\p_t \phi'(t,x)=\p^2 \phi'(t,x)$. However, this equation is not gauge-invariant.
A key point in the GF-ERG equation is to equip a gauge- (or BRST-) invariant form of diffusion equations for an Exact Renormalization Group equation.

Field ingredients in the Wilsonian effective action for sQED are the photon (U(1) gauge), Nakanishi-Lautrup, ghost, anti-ghost, and scalar matter fields.
These fields obey a gauge-covariant version of the diffusion equation in GF-ERG.
This approach allows us to obtain the renormalization group flow consistent with the off-shell BRST transformations for finite cutoff scales.
Then, we have derived the flow equation for the Wilsonian effective action, termed the ``GF-ERG equation."
In addition, we have explicitly written down the conditions for the modified BRST invariance and showed that this is reduced to the ordinary Ward-Takahashi identity under certain assumptions.

For dealing with the GF-ERG equation in sQED, we have performed perturbative calculations around the Gaussian fixed point (free theory) up to the second order of the gauge coupling $e_\tau$.
The zeroth order action, that is, the free part was initiated from the Gaussian fixed point, and the first order correction $O(e_\tau)$ induces a three-point photon-matter-matter vertex.
For second-order contributions $O(e_\tau^2)$, we have studied the photon-photon-matter-matter vertex and the photon two-point correlation function to obtain the anomalous dimension of the photon field. The result is consistent with the standard perturbation theory in four spacetime dimensions.
In addition, our GF-ERG equation yields no one-loop contribution to the photon mass term.
This result is unlike conventional approaches, such as the Wilson-Polchinski equation, which generally induces an artificial photon mass due to the introduction of the cutoff scale. 

Our work has several implications for gauge invariance. In particular, the vertex functions (the $n$-point correlation functions)  satisfy the Ward-Takahashi identity in terms of momenta at each order of the gauge coupling.
This property suggests that the GF-ERG approach could provide a more robust foundation for realizing the gauge-invariant renormalization group flow.
This noteworthy feature could occur not only in sQED but also in broader contexts.

As a potential avenue for future research, the consistency with the standard perturbation theory for higher vertices remains to be examined.
In fact, for the matter sector, its anomalous dimension in QED has been reported to differ from the perturbation theory~\cite{Miyakawa:2021wus}.
Thus, exploring the anomalous dimensions and beta functions of matter interactions in sQED is a significant challenge.

It is also intriguing to explore the relationships between GF-ERG and other formulations, such as those referenced in Refs.~\cite{Morris:1999px,Morris:2000fs,Arnone:2005vd,Rosten:2008zp}.
A critical element in this regard is the coarse-graining operator $\Psi$ in the Wegner~\cref{eq: Wegner equation}.
We have detailed the specific form of $\Psi$ for GF-ERG in \Cref{eqapp: psi operator}.
In a manifestly gauge-invariant ERG formulation, discerning the form of $\Psi$ might be more straightforward.
Comparing $\Psi$ across these approaches could shed light on their differences and similarities.

Furthermore, extensions of GF-ERG to other gauge theories have significant potential. In the context of the asymptotic safety program, which aims to identify UV fixed points in quantum field theories with gravity, GF-ERG could be a vital computational tool for future investigations.

\subsection*{Acknowledgements}
The authors thank Jan M. Pawlowski for intensive discussions at the initial stage of this work and for carefully reading our manuscript.
J.\,H. acknowledges the Institute for Theoretical Physics, Heidelberg University, for the very kind hospitality during his stay.
The authors also thank Hiroshi Suzuki for valuable discussions and comments.
In particular, the proof of the vanishing 1-loop contribution to the photon mass is inspired by the discussion in the QED case with him.
J.\,H. thanks the members of the Elementary Particle Theory Group at Kyushu University for their hospitality during his stay.
The work of J.\,H. is partially supported by JSPS Grant-in-Aid for Scientific Research KAKENHI Grant No. JP21J14825.
The work of M.\,Y. is supported by the National Science Foundation of China (NSFC) under Grant No.\,12205116 and the Seeds Funding of Jilin University.

\appendix

\section{Notation}
We work on the $D$-dimensional Euclidean space.
$\int_x$ and $\int_p$ are defined as
\begin{align}
    \int_x &\coloneqq \int_{\mathbb{R}^D} \df^Dx,\qquad 
    \int_p \coloneqq \int_{\mathbb{R}^D}
    \frac{\df^Dp}{(2\pi)^d}\,.
\end{align}

Fourier transformation of the fields and functional derivatives are defined as
\begin{align}
\phi(x) &\eqqcolon \int_p e^{ipx} \phi(p),\qquad 
\fdv{\phi(p)} \coloneqq \int_x e^{ipx} \fdv{\phi(x)}\,, \\
\phi^*(x) &\eqqcolon \int_p e^{-ipx} \phi^*(p), \qquad \fdv{\phi^*(p)} \coloneqq \int_x e^{-ipx} \fdv{\phi^*(x)}\,.
\end{align}


\section{Explicit form of GF-ERG equation}
\label{app: GF-ERG}
In this section, we discuss the explicit form of GF-ERG equation and the corresponding Wegner equation.
First, we show the explicit form of the GF-ERG equation for sQED. It is given by
\begin{multline}
\label{e:GF-ERGeq explicit}
    \del_\tau e^{S_\tau}
     = \int_k  \biggl\{
     \qty(2k^2 + \frac{D+2}{2} - \gamma_A + k\cdot \del_k)A_\mu(k) \cdot \fdv{A_\mu(k)} +(2k^2+1-\gamma_A)\fdv{A_\mu(-k)}\fdv{A_\mu(k)}
     \\
     + \qty(2k^2+\frac{D}{2} + \gamma_A + k\cdot \del_k)B(k) \cdot \fdv{B(k)}  - (2k^2+\gamma_A)\fdv{B(-k)}\fdv{B(k)}
      \\
     +\qty(2k^2 + \frac{D+4}{2} - \gamma_A +  k\cdot \del_k) c(k) \cdot \fdv{c(x)}
     + \qty(2k^2+\frac{D}{2} + \gamma_A +  k\cdot \del_k) \bar{c}(k) \cdot \fdv{\bar{c}(k)}
     \\ 
     + 2(2k^2+1) \fdv{\bar{c}(-k)} \fdv{c(k)} 
     \\
   +
   \left(2k^2+\frac{D-2}{2}-\gamma_{\phi}+k\cdot\partial_k\right)\phi(k) \cdot \fdv{\phi(k)} 
   +\left(
   2k^2+\frac{D-2}{2}-\gamma_{\phi}+k\cdot\partial_k\right)\phi^*(k) 
  \cdot\fdv{\phi^*(k)}
   \\
   +2\left(2k^2+1-2\gamma_{\phi}\right) 
   \fdv{\phi^*(k)}\fdv{\phi(k)}
   \Biggr\} e^{S_\tau}
\\
   +\int_x\,
   \fdv{\phi^*(x)} \Biggl[
   4ie_\tau\left(
   A_\mu(x)+\fdv{A_\mu(x)}\right)
   \partial_\mu
   -2e_\tau^2
   \left(A_\mu(x)+\fdv{A_\mu(x)}\right)
   \left(A_\mu(x)+\fdv{A_\mu(x)}\right)
   \Biggr]
   \left(\phi^*(x)+\fdv{\phi(x)}\right)e^{S_\tau}
\\
+\int_x\,
   \fdv{\phi(x)} \Biggl[
   -4ie_\tau\left(
   A_\mu(x)+\fdv{A_\mu(x)}\right)
   \partial_\mu
   -2e_\tau^2
   \left(A_\mu(x)+\fdv{A_\mu(x)}\right)
   \left(A_\mu(x)+\fdv{A_\mu(x)}\right)
   \Biggr]
   \left(\phi(x)+\fdv{\phi^*(x)}\right)e^{S_\tau}.
\end{multline}
The last two terms on the right-hand side of this equation originate from the modification of the diffusion equation \eqref{eq: GF diffusion equations}.

Next, let us discuss the relationship between the GF-ERG equation and the Wegner equation.
For notational convention, let us denote the field contents ($A_\mu, B, c, \bar{c}$ and $\phi$) and the corresponding gradient flow equation \eqref{eq: GF diffusion equations} as $\psi_i(x)$ and $F_i[\psi(x)]$, respectively.
That is, the gradient flow equations are rewritten as follows:
\begin{align}
    \del_t \psi_i(t,x) = F_i[\psi(t,x)].
\end{align}
Then, following \cite{Sonoda:2020vut}, we can express the GF-ERG equation as follows:
\begin{align}
    \del_\tau e^{S_\tau[\psi]} 
    = \int_x \fdv{\psi_i(x)}
    \hat{s}^{-1} 
    \qty(
        [ - d_\psi - \gamma_\psi - x\cdot \del_{x} ]\psi_i(x)  -2 F_i[\psi(x)]
    ) \hat{s} e^{S_\tau} .
\end{align}
Therefore, the coarse-graining operator $\Psi$ in the Wegner equation \eqref{eq: Wegner equation} is given by
\begin{align}
    \Psi = \hat{s}^{-1} 
    \qty(
        [ - d_\psi - \gamma_\psi - x\cdot \del_{x} ]\psi_i(x)  -2 F_i[\psi(x)]
    ) 
    \hat{s}\,.
    \label{eqapp: psi operator}
\end{align}
Indeed, the above derivation can be applied to cases with general fields, scrambler operators, and gradient flow equations, under the present definition of the Wilsonian effective action \eqref{eq: renormalization transformation for sQED}.

\section{Concrete form of \texorpdfstring{$X_{\mu\nu}$}{}}
\label{a:Xmunu}
We give the concrete form of $X_{\mu\nu}$, defined by \cref{e:DiffEqXmunu}.
Using the formula in Appendix A in \cite{Miyakawa:2021wus}, we get 
\begin{multline}
    X_{\mu\nu}(p,k,l)=
    \int_0^1 \df \alpha\, 
    \biggl(
	(p+k+l)_\nu V_\mu(\alpha p,\alpha k) e^{\alpha^2(p+k)^2-(p+k+l)^2-l^2}
 +
	p_\mu V_\nu(\alpha (p+k),\alpha l) e^{\alpha^2 (p+k)^2 - p^2 - k^2}
 \biggr)\,.
\end{multline}
By substituting the definition \eqref{e:Vmu} of $(S_1^{\Phi^*\Phi \mc{A}}(p,k))_\mu$ and executing the interal over $\alpha$, we find that $X_{\mu\nu}(p,k,l)$ is given by
\begin{align}
\label{e:Xmunu}
&X_{\mu\nu}(p,k,l) 
 = (p+k+l)_\nu 
\Biggl[
	\frac{(2p+k)_\mu}{2} F\qty((p+k)^2 - (p+k+l)^2 - l^2)
	\nn
&\quad +
	\frac{p^2(p+k)_\mu}{p^2-(p+k)^2-k^2} \biggl(
		F\qty(p^2-(p+k)^2 - k^2 + (p+k)^2 - (p+k+l)^2 - l^2) 
		- F\qty((p+k)^2 - (p+k+l)^2 - l^2)
	\biggr)
 \nn
 &\quad+
	\frac{(p+k)^2p_\mu}{(p+k)^2-p^2-k^2} 
	\biggl(
		F\qty((p+k)^2 - p^2 - k^2 + (p+k)^2 - (p+k+l)^2 - l^2) 
		- F\qty((p+k)^2 - (p+k+l)^2 - l^2)
	\biggr)
\Biggr]
\nn
&\quad 
+ p_\mu 
\Biggl[
	\frac{(2p+2k+l)_\nu}{2} F\qty((p+k)^2 - p^2 - k^2)
	+
	\frac{(p+k)^2 (p+k+l)_\nu}{(p+k)^2-(p+k+l)^2-l^2}
 \nn
 &\qquad\qquad
 \times 
	\biggl(
		F\qty((p+k)^2-(p+k+l)^2 - l^2 + (p+k)^2 - p^2 - k^2) 
		- F\qty((p+k)^2 - p^2 - k^2)
	\biggr)
\nn
&\quad
+
\frac{(p+k+l)^2(p+k)_\nu}{(p+k+l)^2-(p+k)^2-l^2}
	\biggl(
		F\qty((p+k+l)^2-(p+k)^2 - l^2 + (p+k)^2 - p^2 - k^2) 
		- F\qty((p+k)^2 - p^2 - k^2)
	\biggr)
\Biggr]\,.
\end{align}

\section{Modified BRST invariance}
\label{app: Modified BRST invariance}

\subsection{Reduction of modified BRST invariance}
In this section, we show that the modified BRST invariance condition is reduced to the ordinary Ward-Takahashi identity under some assumption.

Before the proof, let us remind that the modified BRST invariance condition with its generator $\tilde{\hat{\mc{G}}}_B$ \eqref{eq: modified BRST generator} is given by
\begin{multline} 
 0  = e^{-S_\tau} \tilde{\hat{\mc{G}}}_B e^{S_\tau}
   = \int_x \biggl[ \del_\mu \qty(c(x) + \fdv{S_\tau}{\bar{c}(x)}) \fdv{S_\tau}{A_\mu(x)} + \qty(B(x) - \fdv{S_\tau}{B(x)}) \fdv{S_\tau}{\bar{c}(x)}
    \\ + \fdv{A_\mu(x)}\del_\mu\fdv{S_\tau}{\bar{c}(x)}  - \sfdv{S_\tau}{B(x)}{\bar{c}(x)} + ie_\tau c(x)\phi(x)\fdv{S_\tau}{\phi(x)} - ie_\tau c(x)\phi^*(x)\fdv{S_\tau}{\phi^*(x)}
    \biggr].
    \label{eq: modified BRST invariance}
\end{multline}

In the following, we decompose $S_\tau$ into the Gaussian part $S_0$ and the interaction part $S_I$ as 
\begin{align}
S_\tau = S_0 + S_I
\end{align}
and show that the above equation is reduced to the ordinary Ward-Takahashi identity.
We assume that $S_I$ is a functional only of $\mathcal{A}_\mu, \mc{B}, \phi$ and $\phi^*$;
\begin{align}
S_I = S_I[\mc{A}_\mu,\mc{B},\phi,\phi^*].
\end{align}
This is true for the case where the ghost sector is free, while the gauge and matter sectors have any interaction.
Substituting this assumption into \Cref{eq: modified BRST invariance} and moving onto the momentum space, we get 
\begin{multline}
   0 = \int_k \biggl[
    i k_\mu  e^{-k^2} C(k) \qty(\fdv{S_0}{A_\mu(k)} +\fdv{S_I}{A_\mu(k)}) + \qty(e^{-k^2} \mathcal{B}(k) - \fdv{S_I}{B(-k)}) \fdv{S_0}{\bar{c}(k)}
    \biggl]
     \\
     +ie_\tau \int_{p,k} \biggl[
     e^{-k^2}  C(k) \phi(p)\qty(\fdv{S_0}{\phi(p+k)} + \fdv{S_I}{\phi(p+k)}) - e^{-k^2} C(k) \phi^*(p+k)\qty(\fdv{S_0}{\phi(p)} + \fdv{S_I}{\phi^*(p)})
    \biggr].
\end{multline}
Using the fact that the Gaussian part $S_0$ saturates the modified BRST invariance condition, that is, this equation also holds for the case with $S_I=0$, we get
\begin{align}
   0 = \int_k \biggl[ik_\mu  e^{-k^2} C(k) \fdv{S_I}{A_\mu(k)} + k^2 e^{k^2} C(k) \fdv{S_I}{B(k)}
     + ie_\tau \int_{p,k}  e^{-k^2}  C(k) \qty( \phi(p)\fdv{S_I}{\phi(p+k)} - \phi^*(p+k)\fdv{S_I}{\phi^*(p)})
    \biggr].
 \end{align}
 Focusing on the coefficients of $ e^{-k^2} C(k)$, we get
  \begin{align}
   0 = ik_\mu \fdv{S_I}{A_\mu(k)} + k^2 e^{2k^2} \fdv{S_I}{B(k)}
     + ie_\tau \int_{p}\qty( \phi(p)\fdv{S_I}{\phi(p+k)} - \phi^*(p+k)\fdv{S_I}{\phi^*(p)}).
     \label{eq: modified BRST SI}
 \end{align}
Let us rewrite the functional derivatives $\delta/\delta A_\mu$ and $\delta/\delta B$ with respect to the $-1$ variables $\mc{A}_\mu$ and $\mc{B}$.
The functional derivatives can be calculated as
\begin{align}
i k_\mu \fdv{S_I}{A_\mu(k)}
& = i k_\mu  \int_p \qty[ \fdv{\mc{A}_\nu(p)}{A_\mu(k)}\fdv{S_I}{\mc{A}_\nu(p)} 
+ \fdv{\mc{B}(p)}{A_\mu(k)}\fdv{S_I}{\mc{B}(p)}] \\
& = i e^{-k^2}k_\mu  h_{\mu\nu}(k) \fdv{S_I}{\mc{A}_\nu(k)} 
- k^2 e^{-k^2}h_B(k^2) \fdv{S_I}{\mc{B}(p)}
\\
& = i e^{-k^2}(\xi + e^{-2k^2 }) h_B(k^2) k_\mu \fdv{S_I}{\mc{A}_\mu(k)} 
- k^2 e^{-k^2}h_B(k^2) \fdv{S_I}{\mc{B}(p)},\\
k^2 e^{2k^2} \fdv{S_I}{B(k)}
& = k^2 e^{2k^2} \int_p 
\qty[
\fdv{\mc{A}_\mu(p)}{B(k)}\fdv{S_I}{\mc{A}_\mu(p)} + \fdv{\mc{B}(p)}{B(k)}\fdv{S_I}{\mc{B}(p)}
] \\
& = i k^2 e^{k^2}h_B(k^2) k_\mu \fdv{S_I}{\mc{A}_\mu(k)} + k^2 e^{-k^2} h_B(k^2) \fdv{S_I}{\mc{B}(k)}.
\end{align}
Summing up these terms, we obtain
\begin{align}
    ik_\mu \fdv{S_I}{A_\mu(k)} + k^2 e^{2k^2} \fdv{S_I}{B(k)} = i e^{k^2} (k^2 + \xi e^{-2k^2} + e^{-4k^2})h_B(k^2) k_\mu \fdv{S_I}{\mc{A}_\mu(k)}
    = ie^{k^2} k_\mu \fdv{S_I}{\mc{A}_\mu(k)}.
\end{align}
Substituting this relation into \Cref{eq: modified BRST SI},
we find that the modified BRST invariance condition for $S_I$ is given by
\begin{align}
   0 = ie^{k^2} k_\mu \fdv{S_I}{\mc{A}_\mu(k)}
     + ie_\tau \int_{p}\qty( \phi(p)\fdv{S_I}{\phi(p+k)} - \phi^*(p+k)\fdv{S_I}{\phi^*(p)}),
 \end{align}
 which is nothing but the ordinary WT identity with $e^{-k^2}\mc{A}_\mu, \phi$ and $\phi^*$.

\subsection{Ward-Takahashi identity for vertices}
In this section, we see how the Ward-Takahashi identity constrains the vertices of the Wilsonian effective action.

\subsubsection*{First order \texorpdfstring{$(S_1^{\Phi^*\Phi\mc{A}})$}{}}
\begin{figure}[hbtp]
\centering
\includegraphics[width=0.5\linewidth,pagebox=cropbox,clip]{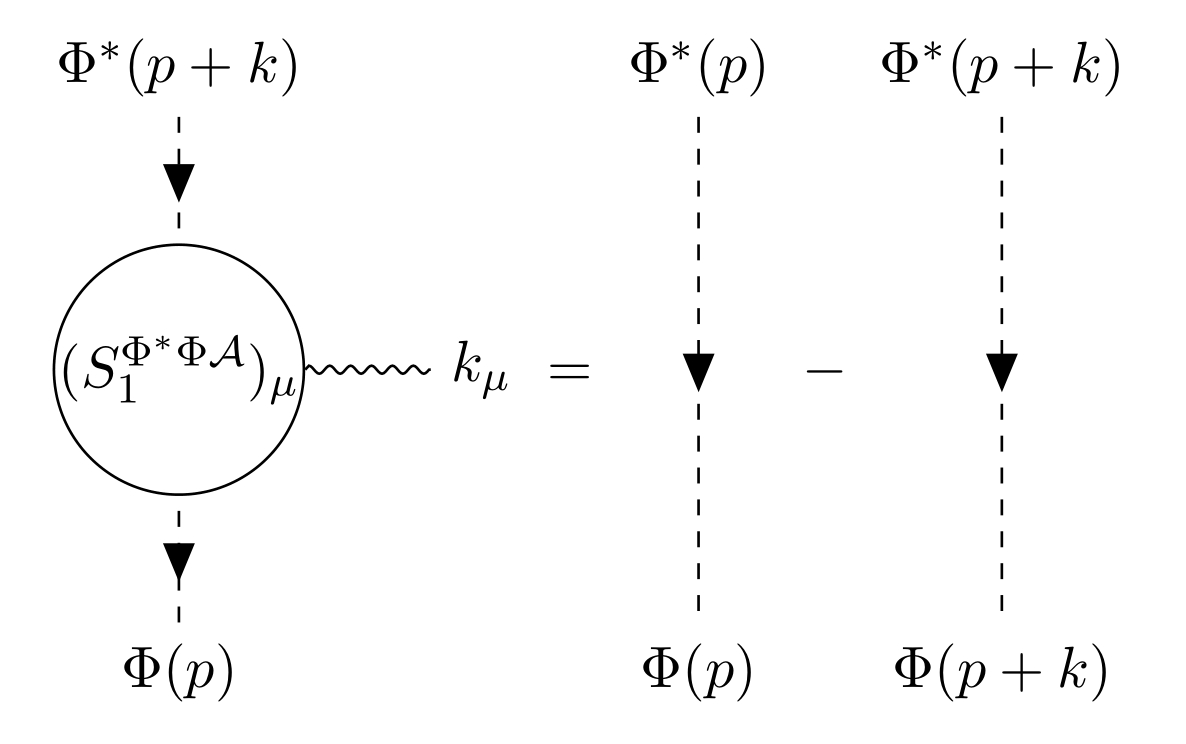}
\caption{Diagrammatic representation of the Ward-Takahashi identity for $\mc{S}_1^{\Phi^*\Phi \mc{A}}$.}
\label{fig: WT identity for three point vertex}
\end{figure}

Let us see how the first order correction $S_1^{\Phi^*\Phi \mc{A}}$ is constrained by the Ward-Takahashi identity.
Substituting $S=\mc{S}^2_0 + e_\tau \mc{S}_1^{\Phi^*\Phi \mc{A}} + \order{e_\tau^2}$ into the WT identity \eqref{e:rWTid} and focusing on $\Phi^*\Phi$-term of $\order{e_\tau}$, we get
\begin{align}
    k_\mu e^{k^2} \fdv{\mc{S}_1^{\Phi^*\Phi \mc{A}}}{\mc{A}_\mu(k)} =
    \int_{p}\qty( \phi^*(p+k)\fdv{\mc{S}^2_0}{\phi^*(p)} - \phi(p)\fdv{\mc{S}^2_0}{\phi(p+k)}),
\end{align}
which is depicted in \Cref{fig: WT identity for three point vertex}.
Substituting the ansatz of $S_1^{\Phi^*\Phi\mc{A}}$ and \Cref{e:VmuWithAB}, 
\begin{align}
\label{e:WTidVmu}
0 &= k_\mu (S_1^{\Phi^*\Phi\mc{A}})_\mu(p,k) -  e^{(p+k)^2-p^2 - k^2} h_S^{-1}(p+k) + e^{p^2-(p+k)^2 - k^2} h_S^{-1}(p)\\
& = (a -2) p\cdot k + (b-1) k^2.
\end{align}
Then, we find that $(a,b)=(2,1)$ satisfies this identity for any $p$ and $k$.

\subsubsection*{Second order \texorpdfstring{$(S_2^{\Phi^*\Phi\mc{A}\mc{A}})$}{}}
\begin{figure}[hbtp]
\centering
\includegraphics[width=0.9\linewidth,pagebox=cropbox,clip]{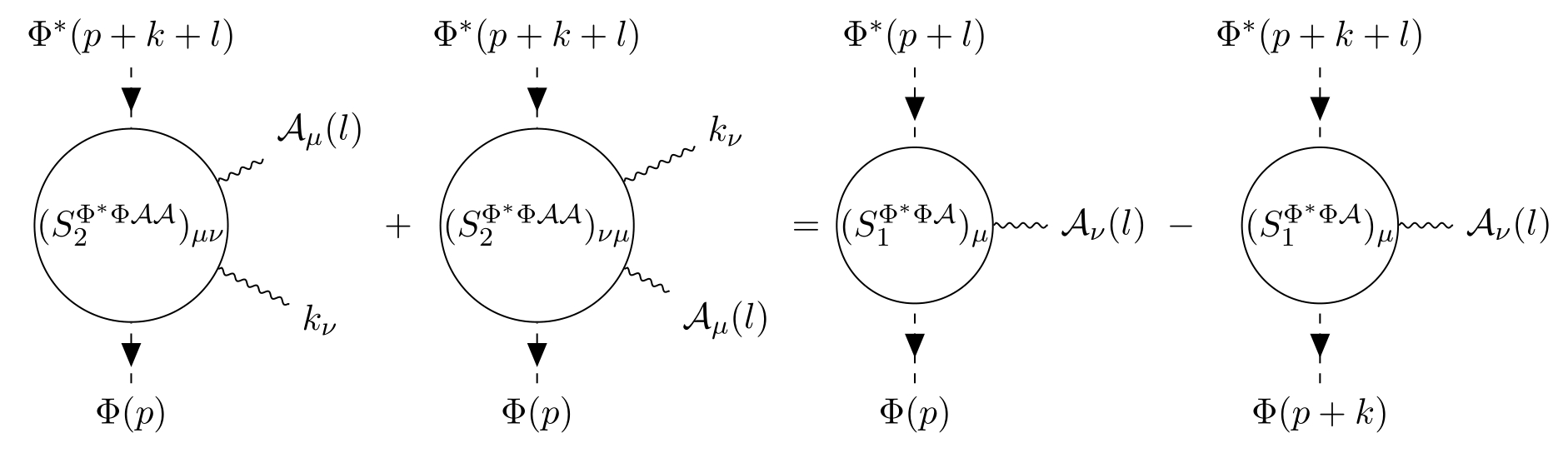}
\caption{Diagrammatic representation of the Ward-Takahashi identity for $\mc{S}_2^{\Phi^*\Phi \mc{A}\mc{A}}$.}
\label{fig: WT identity for four point vertex}
\end{figure}

Let us see how the second order correction to the four-point vertex $S_2^{\Phi^*\Phi \mc{A}\mc{A}}$ is constrained by the Ward-Takahashi identity.
Substituting $S=\mc{S}^2_0 + e_\tau \mc{S}_1^{\Phi^*\Phi \mc{A}} + e_\tau^2 (S_2^{\Phi^*\Phi \mc{A}\mc{A}} +S_2^{\mc{A}\mc{A}}) + \order{e_\tau^3}$ into the WT identity \eqref{e:rWTid} and focusing on $\Phi^*\Phi \mc{A}$-term of $\order{e_\tau^2}$, we get
\begin{align}
    k_\mu e^{k^2} \fdv{\mc{S}_2^{\Phi^*\Phi \mc{A}\mc{A}}}{\mc{A}_\mu(k)} =
    \int_{p}\qty( \phi^*(p+k)\fdv{\mc{S}^{\Phi^*\Phi \mc{A}}_1}{\phi^*(p)} - \phi(p)\fdv{\mc{S}^{\Phi^*\Phi \mc{A}}_1}{\phi(p+k)}),
\end{align}
which is depicted in \Cref{fig: WT identity for four point vertex}.
Substituting the concrete forms of $S_1^{\Phi^* \Phi\mc{A}}$ and $S_2^{\Phi^* \Phi\mc{A}\mc{A}}$, we get
\begin{align}
& 
\begin{multlined}
    0 = k_\mu \qty((S_2^{\Phi^* \Phi\mc{A}\mc{A}})_{\mu\nu}(p,k,l)+(S_2^{\Phi^* \Phi\mc{A}\mc{A}})_{\nu\mu}(p,l,k)) 
    \\
    - 
    e^{(p+k+l)^2-(p+l)^2 - k^2}\frac{h_S(p+l)}{h_S(p+k+l)}(S_1^{\Phi^* \Phi\mc{A}})_\nu(p,l) - e^{p^2-(p+k)^2 - k^2}\frac{h_S(p+k)}{h_S(p)} (S_1^{\Phi^* \Phi\mc{A}})_\nu(p+k,l) 
\end{multlined}
\\ & \phantom{0} = 2 (c+1) k_\nu 
\end{align}
Therefore, we find that the WT identity requires $c$ to be $-1$.

\subsubsection*{Second order \texorpdfstring{$(S_2^{\mc{A}\mc{A}})$}{}}
\begin{figure}[hbtp]
\centering
\includegraphics[width=0.4\linewidth,pagebox=cropbox,clip]{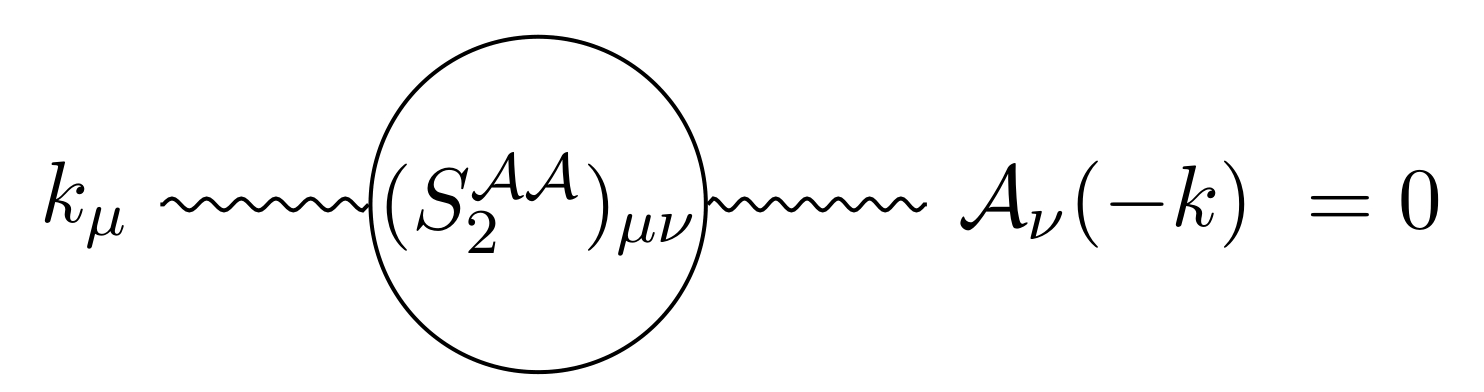}
\caption{Diagrammatic representation of the Ward-Takahashi identity for $\mc{S}_2^{\mc{A}\mc{A}}$.}
\label{fig: WT identity for two point vertex}
\end{figure}

Let us see how the second order correction to the four-point vertex $S_2^{\Phi^*\Phi \mc{A}\mc{A}}$ is constrained by the Ward-Takahashi identity.
Substituting $S=\mc{S}^2_0 + e_\tau \mc{S}_1^{\Phi^*\Phi \mc{A}} + e_\tau^2 (S_2^{\Phi^*\Phi \mc{A}\mc{A}} +S_2^{\mc{A}\mc{A}}) + \order{e_\tau^3}$ into the WT identity \eqref{e:rWTid} and focusing on $\mc{A}$-term of $\order{e_\tau^2}$, we get
\begin{align}
    k_\mu \fdv{\mc{S}_2^{\mc{A}\mc{A}}}{\mc{A}_\mu(k)} = 0,
\end{align}
which is depicted in \Cref{fig: WT identity for two point vertex}.

Substituting the ansatz of $S_2^{\mc{A}\mc{A}}$, we get
\begin{align}
    k_\mu (S_2^{\mc{A}\mc{A}})_{\mu\nu}(k) = 0,
\end{align}
which yields the transversality ($(S_2^{\mc{A}\mc{A}})_{\mu\nu}(k) \propto \delta_{\mu\nu} - k_\mu k_\nu/k^2$) of the two-point function.

\bibliographystyle{JHEP}
\bibliography{ref}

\providecommand{\href}[2]{#2}\begingroup\raggedright\begin{thebibliography}{10}

\bibitem{Wilson:1973jj}
K.~G. Wilson and J.~B. Kogut, \emph{{The Renormalization group and the epsilon
  expansion}}, \href{https://doi.org/10.1016/0370-1573(74)90023-4}{\emph{Phys.
  Rept.} {\bfseries 12} (1974) 75}.

\bibitem{Polchinski:1983gv}
J.~Polchinski, \emph{{Renormalization and Effective Lagrangians}},
  \href{https://doi.org/10.1016/0550-3213(84)90287-6}{\emph{Nucl. Phys. B}
  {\bfseries 231} (1984) 269}.

\bibitem{Wegner:1972ih}
F.~J. Wegner and A.~Houghton, \emph{{Renormalization group equation for
  critical phenomena}},
  \href{https://doi.org/10.1103/PhysRevA.8.401}{\emph{Phys. Rev. A} {\bfseries
  8} (1973) 401}.

\bibitem{Wetterich:1992yh}
C.~Wetterich, \emph{{Exact evolution equation for the effective potential}},
  \href{https://doi.org/10.1016/0370-2693(93)90726-X}{\emph{Phys. Lett. B}
  {\bfseries 301} (1993) 90}
  [\href{https://arxiv.org/abs/1710.05815}{{\ttfamily 1710.05815}}].

\bibitem{Morris:1993qb}
T.~R. Morris, \emph{{The Exact renormalization group and approximate
  solutions}}, \href{https://doi.org/10.1142/S0217751X94000972}{\emph{Int. J.
  Mod. Phys.} {\bfseries A9} (1994) 2411}
  [\href{https://arxiv.org/abs/hep-ph/9308265}{{\ttfamily hep-ph/9308265}}].

\bibitem{Morris:1998da}
T.~R. Morris, \emph{{Elements of the continuous renormalization group}},
  \href{https://doi.org/10.1143/PTPS.131.395}{\emph{Prog. Theor. Phys. Suppl.}
  {\bfseries 131} (1998) 395}
  [\href{https://arxiv.org/abs/hep-th/9802039}{{\ttfamily hep-th/9802039}}].

\bibitem{Berges:2000ew}
J.~Berges, N.~Tetradis and C.~Wetterich, \emph{{Nonperturbative renormalization
  flow in quantum field theory and statistical physics}},
  \href{https://doi.org/10.1016/S0370-1573(01)00098-9}{\emph{Phys. Rept.}
  {\bfseries 363} (2002) 223}
  [\href{https://arxiv.org/abs/hep-ph/0005122}{{\ttfamily hep-ph/0005122}}].

\bibitem{Aoki:2000wm}
K.~Aoki, \emph{{Introduction to the nonperturbative renormalization group and
  its recent applications}},
  \href{https://doi.org/10.1016/S0217-9792(00)00092-3}{\emph{Int.J.Mod.Phys.}
  {\bfseries B14} (2000) 1249}.

\bibitem{Bagnuls:2000ae}
C.~Bagnuls and C.~Bervillier, \emph{{Exact renormalization group equations. An
  Introductory review}},
  \href{https://doi.org/10.1016/S0370-1573(00)00137-X}{\emph{Phys. Rept.}
  {\bfseries 348} (2001) 91}
  [\href{https://arxiv.org/abs/hep-th/0002034}{{\ttfamily hep-th/0002034}}].

\bibitem{Polonyi:2001se}
J.~Polonyi, \emph{{Lectures on the functional renormalization group method}},
  \href{https://doi.org/10.2478/BF02475552}{\emph{Central Eur. J. Phys.}
  {\bfseries 1} (2003) 1}
  [\href{https://arxiv.org/abs/hep-th/0110026}{{\ttfamily hep-th/0110026}}].

\bibitem{Pawlowski:2005xe}
J.~M. Pawlowski, \emph{{Aspects of the functional renormalisation group}},
  \href{https://doi.org/10.1016/j.aop.2007.01.007}{\emph{Annals Phys.}
  {\bfseries 322} (2007) 2831}
  [\href{https://arxiv.org/abs/hep-th/0512261}{{\ttfamily hep-th/0512261}}].

\bibitem{Gies:2006wv}
H.~Gies, \emph{{Introduction to the functional RG and applications to gauge
  theories}}, \href{https://doi.org/10.1007/978-3-642-27320-9_6}{\emph{Lect.
  Notes Phys.} {\bfseries 852} (2012) 287}
  [\href{https://arxiv.org/abs/hep-ph/0611146}{{\ttfamily hep-ph/0611146}}].

\bibitem{Delamotte:2007pf}
B.~Delamotte, \emph{{An Introduction to the nonperturbative renormalization
  group}}, \href{https://doi.org/10.1007/978-3-642-27320-9_2}{\emph{Lect. Notes
  Phys.} {\bfseries 852} (2012) 49}
  [\href{https://arxiv.org/abs/cond-mat/0702365}{{\ttfamily
  cond-mat/0702365}}].

\bibitem{Rosten:2010vm}
O.~J. Rosten, \emph{{Fundamentals of the Exact Renormalization Group}},
  \href{https://doi.org/10.1016/j.physrep.2011.12.003}{\emph{Phys. Rept.}
  {\bfseries 511} (2012) 177}
  [\href{https://arxiv.org/abs/1003.1366}{{\ttfamily 1003.1366}}].

\bibitem{Kopietz:2010zz}
P.~Kopietz, L.~Bartosch and F.~Sch\"utz, \emph{{Introduction to the functional
  renormalization group}}, vol.~798. Springer Berlin, Heidelberg, 2010,
  \href{https://doi.org/10.1007/978-3-642-05094-7}{10.1007/978-3-642-05094-7}.

\bibitem{Braun:2011pp}
J.~Braun, \emph{{Fermion Interactions and Universal Behavior in Strongly
  Interacting Theories}},
  \href{https://doi.org/10.1088/0954-3899/39/3/033001}{\emph{J. Phys.}
  {\bfseries G39} (2012) 033001}
  [\href{https://arxiv.org/abs/1108.4449}{{\ttfamily 1108.4449}}].

\bibitem{Dupuis:2020fhh}
N.~Dupuis, L.~Canet, A.~Eichhorn, W.~Metzner, J.~M. Pawlowski, M.~Tissier
  et~al., \emph{{The nonperturbative functional renormalization group and its
  applications}},
  \href{https://doi.org/10.1016/j.physrep.2021.01.001}{\emph{Phys. Rept.}
  {\bfseries 910} (2021) 1} [\href{https://arxiv.org/abs/2006.04853}{{\ttfamily
  2006.04853}}].

\bibitem{Litim:1998wk}
D.~F. Litim and J.~M. Pawlowski, \emph{{On gauge invariance and Ward identities
  for the Wilsonian renormalization group}},
  \href{https://doi.org/10.1016/S0920-5632(99)00187-5}{\emph{Nucl. Phys. B
  Proc. Suppl.} {\bfseries 74} (1999) 325}
  [\href{https://arxiv.org/abs/hep-th/9809020}{{\ttfamily hep-th/9809020}}].

\bibitem{Litim:1998nf}
D.~F. Litim and J.~M. Pawlowski, \emph{{On gauge invariant Wilsonian flows}},
  in \emph{{Workshop on the Exact Renormalization Group}}, pp.~168--185, 9,
  1998, \href{https://arxiv.org/abs/hep-th/9901063}{{\ttfamily
  hep-th/9901063}}.

\bibitem{Wetterich:2016ewc}
C.~Wetterich, \emph{{Gauge invariant flow equation}},
  \href{https://doi.org/10.1016/j.nuclphysb.2018.04.020}{\emph{Nucl. Phys. B}
  {\bfseries 931} (2018) 262}
  [\href{https://arxiv.org/abs/1607.02989}{{\ttfamily 1607.02989}}].

\bibitem{Asnafi:2018pre}
S.~Asnafi, H.~Gies and L.~Zambelli, \emph{{BRST invariant RG flows}},
  \href{https://doi.org/10.1103/PhysRevD.99.085009}{\emph{Phys. Rev. D}
  {\bfseries 99} (2019) 085009}
  [\href{https://arxiv.org/abs/1811.03615}{{\ttfamily 1811.03615}}].

\bibitem{Igarashi:2019gkm}
Y.~Igarashi, K.~Itoh and T.~R. Morris, \emph{{BRST in the exact renormalization
  group}}, \href{https://doi.org/10.1093/ptep/ptz099}{\emph{PTEP} {\bfseries
  2019} (2019) 103B01} [\href{https://arxiv.org/abs/1904.08231}{{\ttfamily
  1904.08231}}].

\bibitem{Igarashi:2021zml}
Y.~Igarashi and K.~Itoh, \emph{{QED in the exact renormalization group}},
  \href{https://doi.org/10.1093/ptep/ptab142}{\emph{PTEP} {\bfseries 2021}
  (2021) 123B06} [\href{https://arxiv.org/abs/2107.14012}{{\ttfamily
  2107.14012}}].

\bibitem{Igarashi:2016gcf}
Y.~Igarashi, K.~Itoh and J.~M. Pawlowski, \emph{{Functional flows in QED and
  the modified Ward\textendash{}Takahashi identity}},
  \href{https://doi.org/10.1088/1751-8113/49/40/405401}{\emph{J. Phys. A}
  {\bfseries 49} (2016) 405401}
  [\href{https://arxiv.org/abs/1604.08327}{{\ttfamily 1604.08327}}].

\bibitem{Fejos:2016wza}
G.~Fejos and T.~Hatsuda, \emph{{Fixed point structure of the Abelian Higgs
  model}}, \href{https://doi.org/10.1103/PhysRevD.93.121701}{\emph{Phys. Rev.
  D} {\bfseries 93} (2016) 121701}
  [\href{https://arxiv.org/abs/1604.05849}{{\ttfamily 1604.05849}}].

\bibitem{Fejos:2017sjl}
G.~Fejos and T.~Hatsuda, \emph{{Renormalization group flows of the N-component
  Abelian Higgs model}},
  \href{https://doi.org/10.1103/PhysRevD.96.056018}{\emph{Phys. Rev. D}
  {\bfseries 96} (2017) 056018}
  [\href{https://arxiv.org/abs/1705.07333}{{\ttfamily 1705.07333}}].

\bibitem{Morris:1999px}
T.~R. Morris, \emph{{A Gauge invariant exact renormalization group. 1.}},
  \href{https://doi.org/10.1016/S0550-3213(99)00821-4}{\emph{Nucl. Phys. B}
  {\bfseries 573} (2000) 97}
  [\href{https://arxiv.org/abs/hep-th/9910058}{{\ttfamily hep-th/9910058}}].

\bibitem{Morris:2000fs}
T.~R. Morris, \emph{{A Gauge invariant exact renormalization group. 2.}},
  \href{https://doi.org/10.1088/1126-6708/2000/12/012}{\emph{JHEP} {\bfseries
  12} (2000) 012} [\href{https://arxiv.org/abs/hep-th/0006064}{{\ttfamily
  hep-th/0006064}}].

\bibitem{Morris:2005tv}
T.~R. Morris and O.~J. Rosten, \emph{{A Manifestly gauge invariant, continuum
  calculation of the SU(N) Yang-Mills two-loop beta function}},
  \href{https://doi.org/10.1103/PhysRevD.73.065003}{\emph{Phys. Rev. D}
  {\bfseries 73} (2006) 065003}
  [\href{https://arxiv.org/abs/hep-th/0508026}{{\ttfamily hep-th/0508026}}].

\bibitem{Arnone:2005vd}
S.~Arnone, T.~R. Morris and O.~J. Rosten, \emph{{Manifestly gauge invariant
  QED}}, \href{https://doi.org/10.1088/1126-6708/2005/10/115}{\emph{JHEP}
  {\bfseries 10} (2005) 115}
  [\href{https://arxiv.org/abs/hep-th/0505169}{{\ttfamily hep-th/0505169}}].

\bibitem{Rosten:2008zp}
O.~J. Rosten, \emph{{A Resummable beta-Function for Massless QED}},
  \href{https://doi.org/10.1016/j.physletb.2008.03.006}{\emph{Phys. Lett. B}
  {\bfseries 662} (2008) 237}
  [\href{https://arxiv.org/abs/0801.2462}{{\ttfamily 0801.2462}}].

\bibitem{Morris:2016nda}
T.~R. Morris and A.~W.~H. Preston, \emph{{Manifestly diffeomorphism invariant
  classical Exact Renormalization Group}},
  \href{https://doi.org/10.1007/JHEP06(2016)012}{\emph{JHEP} {\bfseries 06}
  (2016) 012} [\href{https://arxiv.org/abs/1602.08993}{{\ttfamily
  1602.08993}}].

\bibitem{Sonoda:2020vut}
H.~Sonoda and H.~Suzuki, \emph{{Gradient flow exact renormalization group}},
  \href{https://doi.org/10.1093/ptep/ptab006}{\emph{PTEP} {\bfseries 2021}
  (2021) 023B05} [\href{https://arxiv.org/abs/2012.03568}{{\ttfamily
  2012.03568}}].

\bibitem{Miyakawa:2021hcx}
Y.~Miyakawa and H.~Suzuki, \emph{{Gradient flow exact renormalization group:
  Inclusion of fermion fields}},
  \href{https://doi.org/10.1093/ptep/ptab100}{\emph{PTEP} {\bfseries 2021}
  (2021) 083B04} [\href{https://arxiv.org/abs/2106.11142}{{\ttfamily
  2106.11142}}].

\bibitem{Miyakawa:2021wus}
Y.~Miyakawa, H.~Sonoda and H.~Suzuki, \emph{{Manifestly gauge invariant exact
  renormalization group for quantum electrodynamics}},
  \href{https://doi.org/10.1093/ptep/ptac003}{\emph{PTEP} {\bfseries 2022}
  (2022) 023B02} [\href{https://arxiv.org/abs/2111.15529}{{\ttfamily
  2111.15529}}].

\bibitem{Abe:2022smm}
Y.~Abe, Y.~Hamada and J.~Haruna, \emph{{Fixed point structure of the gradient
  flow exact renormalization group for scalar field theories}},
  \href{https://doi.org/10.1093/ptep/ptac021}{\emph{PTEP} {\bfseries 2022}
  (2022) 033B03} [\href{https://arxiv.org/abs/2201.04111}{{\ttfamily
  2201.04111}}].

\bibitem{Miyakawa:2023yob}
Y.~Miyakawa, H.~Sonoda and H.~Suzuki, \emph{{Chiral anomaly as a composite
  operator in the gradient flow exact renormalization group formalism}},
  \href{https://doi.org/10.1093/ptep/ptad074}{\emph{PTEP} {\bfseries 2023}
  (2023) 063B03} [\href{https://arxiv.org/abs/2304.14753}{{\ttfamily
  2304.14753}}].

\bibitem{Dutta:2020vqo}
S.~Dutta, B.~Sathiapalan and H.~Sonoda, \emph{{Wilson action for the $O(N)$
  model}}, \href{https://doi.org/10.1016/j.nuclphysb.2020.115022}{\emph{Nucl.
  Phys. B} {\bfseries 956} (2020) 115022}
  [\href{https://arxiv.org/abs/2003.02773}{{\ttfamily 2003.02773}}].

\bibitem{Wegner:1976bn}
F.~J. Wegner, \emph{{The Critical State, General Aspects}},  in \emph{{12th
  School of Modern Physics on Phase Transitions and Critical Phenomena}}, 1976.

\bibitem{Latorre:2000qc}
J.~I. Latorre and T.~R. Morris, \emph{{Exact scheme independence}},
  \href{https://doi.org/10.1088/1126-6708/2000/11/004}{\emph{JHEP} {\bfseries
  11} (2000) 004} [\href{https://arxiv.org/abs/hep-th/0008123}{{\ttfamily
  hep-th/0008123}}].

\end{thebibliography}\endgroup

\end{document}